\newcommand{\Zs}{$Z_\odot\,$}			
\newcommand{\Ms}{$M_\odot\,$}			
\newcommand{\Ho}{$H_{0}\,$}			
\newcommand{\Hp}{$H_{p}\,$}			
\newcommand{\zf}{$z_{f}\,$}			
\newcommand{\logT}{$\log T_{\rm eff}\,$}	
\newcommand{\logL}{$\log L/L_\odot\,$}		
\documentstyle[emulateapj,apjfonts]{article}
\received{MARCH 1999}
\slugcomment{To appear in ApJ (Feb 10, 2000), vol. 530}

\lefthead{Yi et al.}
\righthead{On the Age Estimation of LBDS\,53W091}
 
\begin{document}

\title{ON THE AGE ESTIMATION OF LBDS\,53W091}

\author{Sukyoung Yi}
\affil{Center for Space Astrophysics, Yonsei University, Seoul 120-749, Korea,\\ and California Institute of Technology, Mail Code 405-47, Pasadena, CA 91125\\ yi@srl.caltech.edu}

\author{Tom Brown, Sara Heap, Ivan Hubeny, Wayne Landsman, Thierry Lanz, Allen Sweigart}
\affil{Laboratory for Astronomy and Solar Physics, Code 681, NASA Goddard Space Flight Center, Greenbelt, MD 20771}

\begin{abstract}

	The recent spectral analysis of LBDS\,53W091 by Spinrad and his
collaborators has suggested that this red galaxy at $z$ = 1.552 is at least 
3.5 Gyr old.
        This imposes an important constraint on cosmology, suggesting 
that this galaxy formed at $z \gtrsim 6.5$, assuming recent estimates of
cosmological parameters.
	While their analysis was heavily focused on the use of some UV 
spectral breaks as age indicators, we have performed $\chi^{2}$ tests to 
the continuum of this galaxy using its UV spectrum and photometric data 
($R$, $J$, $H$, \& $K$: 2000 -- 9000 \AA\, in rest-frame).
	We have used the updated Yi models that are based on the Yale tracks.
	We find it extremely difficult to reproduce such large age estimates, 
under the assumption of the most probable input parameters.
	Using the same configuration as in Spinrad et al. (conventional
solar abundance models), our analysis suggests an age of approximately 
1.4 -- 1.8 Gyr.

	We have improved our models over conventional ones by taking 
into account convective core overshoot in the stellar model calculations and
realistic metallicity distributions in the galaxy population synthesis.
	Overshoot affects the visible continuum normalized to the
UV and raises the photometry-based age estimates by 25\%.	
	The use of metallicity mixtures affects the whole spectrum and raises 
all continuum-based age estimates by up to a factor of two.
	If the mean metallicity of the stars in this galaxy is assumed to
be twice solar, the models including these two effects match the UV spectrum 
and photometric data of LBDS\,53W091 near the age of 1.5 -- 2.0 Gyr.
	Our results cannot be easily reconciled with that of Spinrad et al.
(1997).

        The discrepancy between Spinrad et al.'s age estimate (based on the 
Jimenez models) and ours originates from the large difference in the model 
integrated spectrum: the Jimenez models are much bluer than the Yi models and 
the Bruzual \& Charlot models.
         We propose to apply some viable tests to them for verification and search
for the origin of the difference through a more thorough investigation.
	Considering the significance of the age estimates of distant galaxies
as probes of cosmology, it would be an urgent task.

\end{abstract}


\section{Introduction}

	Precise age estimates of high redshift ($z$) galaxies directly
constrain the epoch of galaxy formation, \zf, where \zf is defined 
as the epoch when the majority of stars formed.
	Constraining \zf is important to cosmology.
	For example, one of the key questions in modern cosmology
has been whether the majority of stars in giant elliptical 
galaxies form at high redshifts through violent starbursts or during rather 
recent merger/interaction activities between smaller galaxies.
	In addition, the age of a galaxy is a unique product of just a few 
cosmological parameters (e.g., $\Omega$, $\Lambda$, \Ho, \zf); thus, it can be
used to constrain cosmological parameters as well.

	Spinrad and his collaborators recently obtained the rest-frame UV 
spectrum of LBDS\,53W091, a very red galaxy at $z = 1.552$, using the Keck 
Telescope (Dunlop et al. 1996; Spinrad et al. 1997).
	Based on their analysis on the UV spectrum and $R-K$ color,
they have concluded that LBDS\,53W091 is at least 3.5 Gyr old already 
at $z = 1.552$, which suggests \Ho $<$~45~km~sec$^{-1}$~Mpc$^{-1}$, 
in the context of the Einstein-de Sitter cosmology.
	When more recently measured cosmological parameters are used
(e.g., \Ho = 65, $\Omega = 0.3$, $\Lambda = 0.7$; Aldering et al. 1998),
this age estimate suggests that LBDS 53W091 formed at $z_{f} \gtrsim 6.5$.
 
	Spinrad et al.'s results have been disputed by two independent 
studies. 
	Bruzual \& Magris (1997a) combined the UV spectrum studied by 
Spinrad et al. with $R$, $J$, $H$, $K$ photometry and obtained an age estimate 
1.4 Gyr.
	Heap et al. (1998) interpreted the UV spectral breaks using model 
atmosphere results specifically constructed for their study and using the 
1997 version Yale isochrones with no convective core overshoot 
(see \S 3).
        They have estimated that the age of LBDS\,53W091 lies between 
1 and 2 Gyr.
	They found that the Kurucz spectral library, which is currently 
used in virtually all population synthesis studies including that of Spinrad 
et al., does not match the detailed spectral features in the UV spectrum of 
an F-type main sequence (MS) star obtained with HST/STIS, in terms of the
magnitudes of the spectral breaks used in the analysis of 
Spinrad et al. (1997).
	This implies that Spinrad et al.'s age estimate using UV spectral 
breaks may suffer from some systematic errors.
        Such discordant age estimates undermine our efforts to use this 
important technique as a probe of cosmology.
 
	We have carried out a similar exercise, estimating the age of this 
galaxy, using the Yi population synthesis models 
(Yi, Demarque, \& Oemler 1997).
	In this paper, we present the results from the UV -- visible
continuum analysis only.
	An analysis on the UV spectral breaks is currently in progress.
	We attempt to improve our age estimate by adopting
(1) convective core overshoot in stellar model construction and 
(2) a realistic metallicity distribution, rather than the single-abundance 
assumption, in the galaxy population synthesis.
	We search for the sources of disagreement among various
age estimates and provide a more reliable estimate.

	Our analysis leads to an age estimate for LBDS\,53W091 of approximately
1.5 -- 2.0 Gyr, only half of Spinrad et al.'s estimate, but consistent with 
those of Bruzual \& Magris (1997a) and Heap et al. (1998).
	This smaller age estimate relaxes the strong constraint on \Ho and/or
on \zf implied by the Spinrad et al. estimate.

\section{Model Construction}

	An important advantage of working with the spectra of giant 
elliptical galaxies at $z \approx 1$ -- 2 comes from the fact that their 
major light sources in the UV and visible are all relatively well understood.
	This redshift range corresponds approximately to the age of 1 -- 5 Gyr,
depending on the cosmology adopted.
	Most of the UV light in 1 -- 5 Gyr-old stellar populations 
comes from stars near the main sequence turn-off (MSTO).
	Because the mean temperature of MSTO stars in a coeval stellar 
population is a reliable indicator of the age of the population, the integrated
UV spectra of such galaxies can provide direct clues to their ages.
	At such small ages, in fact, the whole spectrum is a reasonable
age indicator, because the spectral evolution is rapid with time and the 
stellar evolution of all major light sources (the MS and the red giant branch 
[RGB]) is reasonably well understood.
	Thus, we have used both UV and visible data of LBDS\,53W091 
for the age estimation.

	For the purposes of this population synthesis, we have updated 
the Yi models (Yi et al. 1997b) by using isochrones that are more carefully
constructed than before. 
	We have constructed stellar evolutionary tracks for the masses of 
0.4 -- 2.2 \Ms and the metallicities of $Z$ = 0.005, 0.02, and 0.04, using 
the Yale code.
	The tracks have been constructed with up-to-date input physics, 
including the 1995 version OPAL opacities (introduced in 
Rogers \& Iglesias 1992), as described in Yi, Demarque, \& Kim (1997).
	Figure 1 shows a set of stellar evolutionary tracks for $Z = 0.02$
(approximately solar).
	The mean masses of MSTO stars in 1 -- 5 Gyr, solar abundance 
populations range approximately 2.2 -- 1.3 \Ms. 

\placefigure{fig1}
\parbox{3.0in}{\epsfxsize=3.0in \epsfbox{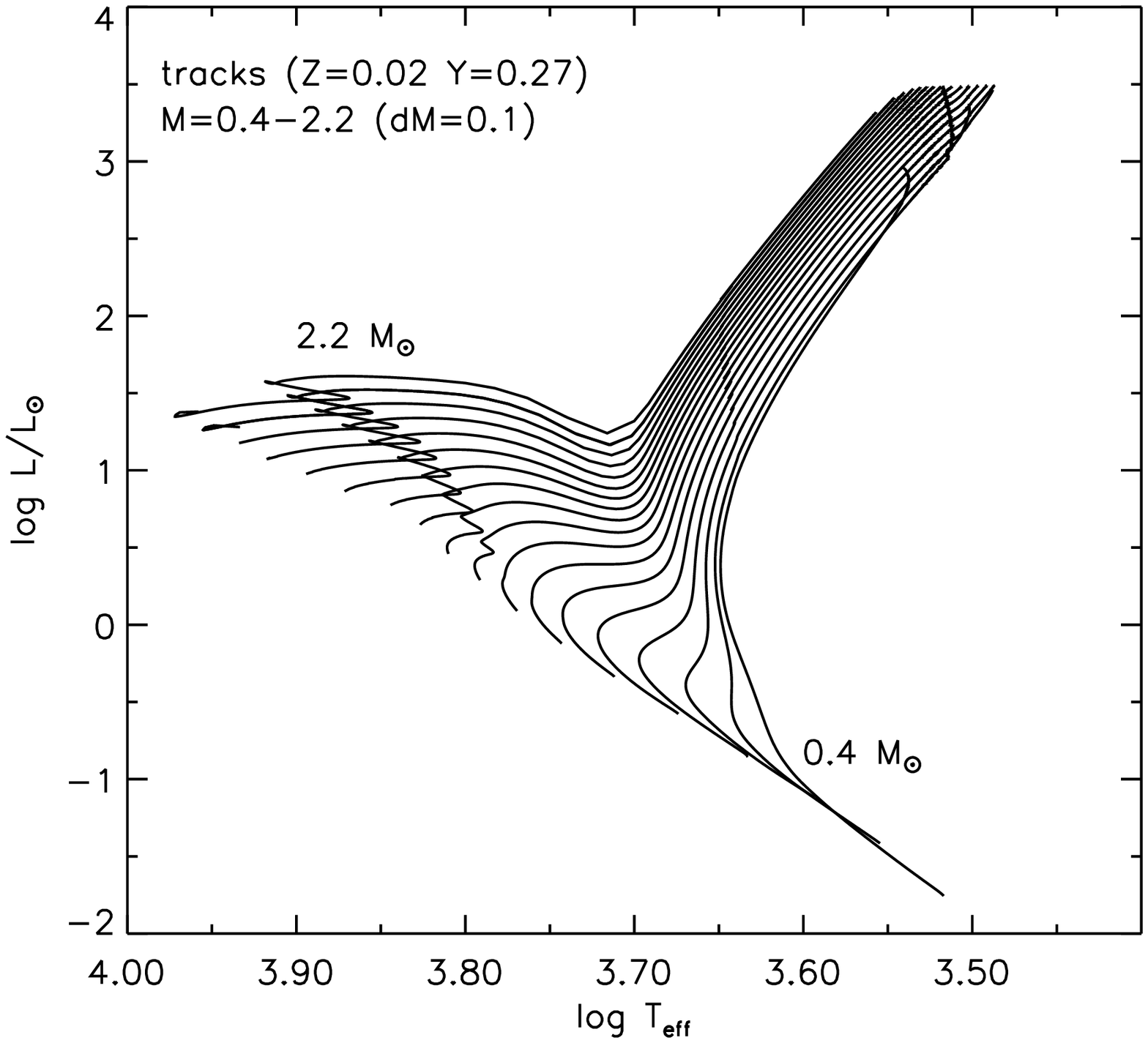}}
\centerline{\parbox{3.0in}{\small {\sc Fig. 1}
Stellar evolutionary tracks from the MS through RGB for Z=0.02, Y=0.27, and
M = 0.4 -- 2.2 \Ms.
}}
\vspace{0.2in}
\addtocounter{figure}{1}

	It is not a simple task to construct isochrones from stellar 
evolutionary tracks, mainly because the interpolation between tracks is not 
trivial.
	Even if the same tracks are used, different isochrone routines may
introduce notable disagreements.	
	The mass interpolation must be carried out with particular caution.
	A small error in mass can cause a seriously inaccurate 
luminosity function near the tip of the RGB, which will in turn affect the 
integrated spectrum.
	For example, when 60 points define one isochrone from the
zero-age MS through RGB, an error in mass in the 4th (when age $<$ 5 Gyr)
-- 5th (age $\gtrsim 5$ Gyr) digit below the decimal (in \Ms) causes a 
noticeable difference in the integrated magnitudes in the near-infrared,
but not necessarily in the shape of the isochrone.
 
	The tip of the RGB is also difficult to locate, partly 
because it is somewhat sensitive to the adopted physics.
	Despite that, it is still important to define it as precisely and 
consistently as possible.
	Because the visible flux is dominated by bright red giants,
an error in  the position of the RGB tip leads to an error in the visible 
flux and thus in the normalized UV flux.
	The UV spectrum is relatively immune againt such complexities
regarding the RGB construction, and thus one may argue that a UV spectrum 
is a better age indicator than the spectrum in longer wavelength regions.
	
\placefigure{fig2}
\parbox{3.0in}{\epsfxsize=3.0in \epsfbox{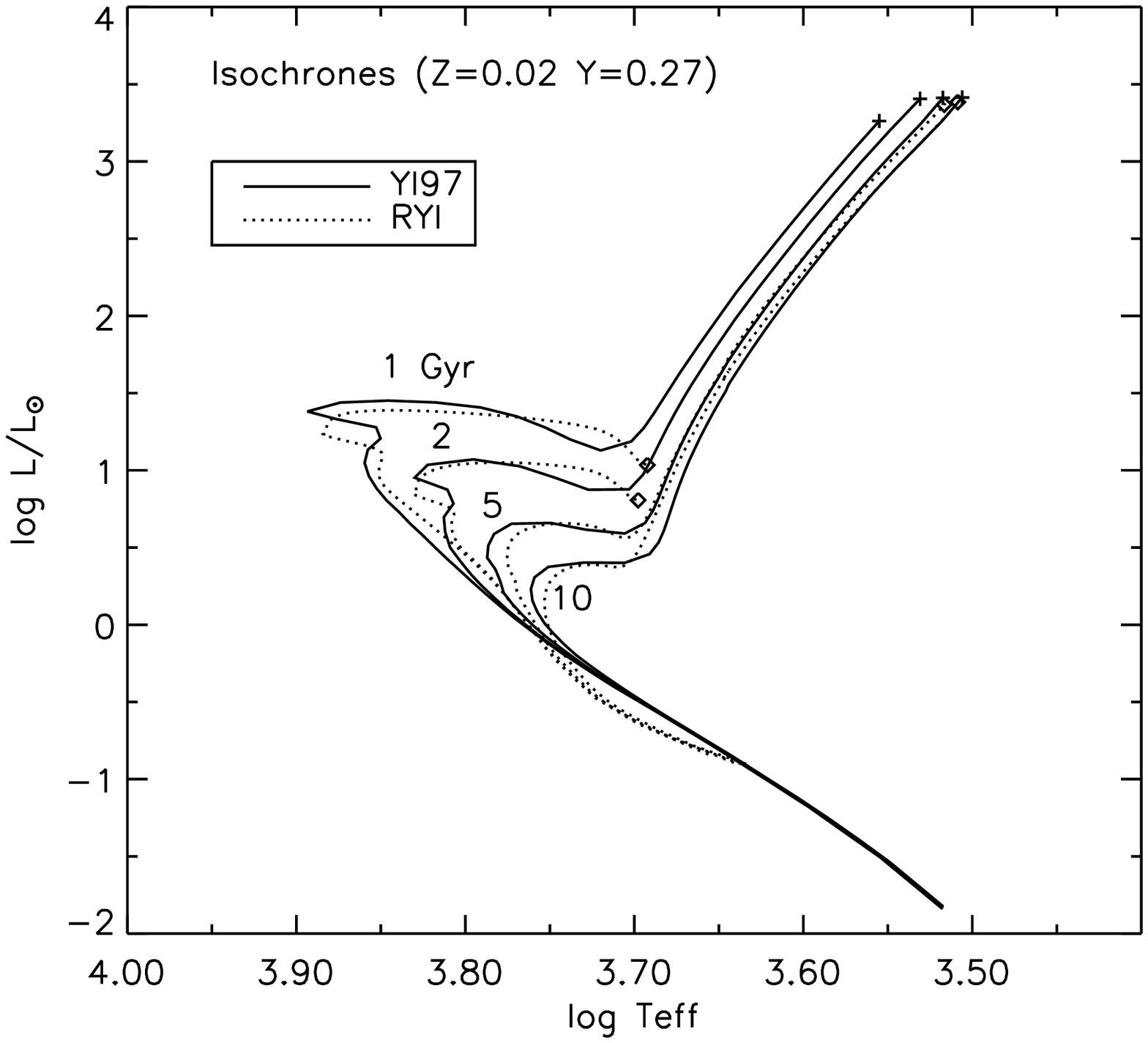}}
\centerline{\parbox{3.0in}{\small {\sc Fig. 2}
New Yale Isochrones for Z=0.02, Y=0.27, t = 1, 2, 5, \& 10 Gyr compared with 
the Revised 
Yale Isochrones (Green et al. 1987). The last point in each isochrone has been
marked with $diamonds$ and $crosses$. It is important to define 
the RGB tips carefully in the UV population synthesis.
}}
\vspace{0.2in}
\addtocounter{figure}{2}

	Figure 2 shows a set of new isochrones constructed from the tracks 
shown in Figure 1, compared with those 
in the Revised Yale Isochrones (RYI; Green, Demarque, \& King 1987).
	The use of improved opacities and energy generation rates has
led to a change in the shape of the isochrones.
	The RYI were designed mainly for isochrone fitting to the
observed color-magnitude diagrams (CMDs) of globular clusters, and, thus,
its creators did not pay much attention to the precise location
of the RGB tips.
	As a result, it is not uncommon to see in the RYI some discontinuities 
in the brightness of RGB tips as a function of age, not to mention that some 
isochrones for small ages do not reach the RGB tip at all.
	The isochrones that are used in this study are shown in Figure 3.
	They are available from this paper through the Astrophysical Journal. 
	A sample is shown in Table 1.
	A complete set of the new Yale Isochrones will be published soon.

\placetable{tbl-1}
\begin{table*}
\caption{Yale Isochornes 1997: a sample for age = 1 Gyr, $Z = 0.02$, $Y=0.27$, 
and OS = 0\tablenotemark{a}} \label{tbl-1}
\begin{center}
\begin{tabular}{ccc}
\tableline
\tableline
Mass (\Ms) & \logT & \logL \\
\tableline
 0.4000000 & 3.5176 & -1.8407\\
 0.4512092 & 3.5360 & -1.6738\\
 0.5031496 & 3.5544 & -1.5073\\
 0.5549585 & 3.5755 & -1.3443\\
 0.6073884 & 3.5966 & -1.1818\\
 0.6617560 & 3.6192 & -1.0226\\
 0.7174781 & 3.6419 & -0.8641\\
 0.7759390 & 3.6651 & -0.7079\\
 0.8362966 & 3.6886 & -0.5529\\
 0.8973278 & 3.7125 & -0.3990\\
 0.8973278 & 3.7125 & -0.3990\\
 0.9292600 & 3.7230 & -0.3257\\
 0.9612948 & 3.7333 & -0.2522\\
 0.9931888 & 3.7436 & -0.1786\\
 1.0279125 & 3.7531 & -0.1043\\
 1.0633708 & 3.7623 & -0.0289\\
 1.0986109 & 3.7713 & 0.0461\\
 1.1368794 & 3.7796 & 0.1219\\
 1.1753701 & 3.7877 & 0.1980\\
 1.2151288 & 3.7956 & 0.2743\\
 1.2575005 & 3.8032 & 0.3505\\
 1.2997627 & 3.8108 & 0.4266\\
 1.3475087 & 3.8181 & 0.5025\\
 1.3959177 & 3.8252 & 0.5790\\
 1.4484234 & 3.8326 & 0.6539\\
 1.5025389 & 3.8393 & 0.7301\\
 1.5604186 & 3.8466 & 0.8048\\
 1.6214427 & 3.8525 & 0.8823\\
 1.6869364 & 3.8571 & 0.9627\\
 1.7595468 & 3.8596 & 1.0476\\
 1.8357059 & 3.8577 & 1.1327\\
 1.9051671 & 3.8502 & 1.2060\\
 1.9590699 & 3.8528 & 1.2785\\
 1.9677832 & 3.8759 & 1.3307\\
 1.9703208 & 3.8935 & 1.3813\\
 1.9929478 & 3.8738 & 1.4393\\
 2.1043307 & 3.8458 & 1.4514\\
 2.1094252 & 3.8175 & 1.4395\\
 2.1129504 & 3.7910 & 1.4071\\
 2.1155551 & 3.7694 & 1.3488\\
 2.1174183 & 3.7525 & 1.2761\\
 2.1190669 & 3.7377 & 1.1991\\
 2.1214807 & 3.7199 & 1.1285\\
 2.1259794 & 3.7018 & 1.1871\\
 2.1291104 & 3.6944 & 1.2742\\
 2.1321655 & 3.6888 & 1.3627\\
 2.1354712 & 3.6836 & 1.4514\\
 2.1390638 & 3.6784 & 1.5400\\
 2.1429518 & 3.6732 & 1.6286\\
 2.1429518 & 3.6732 & 1.6286\\
 2.1514939 & 3.6616 & 1.8154\\
 2.1593514 & 3.6495 & 2.0018\\
 2.1689492 & 3.6398 & 2.1507\\
 2.1730153 & 3.6264 & 2.3366\\
 2.1761145 & 3.6128 & 2.5223\\
 2.1784258 & 3.5989 & 2.7079\\
 2.1801612 & 3.5847 & 2.8932\\
 2.1814793 & 3.5701 & 3.0782\\
 2.1849382 & 3.5551 & 3.2624\\
\tableline
\end{tabular}
\end{center}
\tablenotetext{a}{New Yale Isochrones for other compositions and ages are
available from the Astrophysical Journal.}
\end{table*}

\placefigure{fig3}
\parbox{3.0in}{\epsfxsize=3.0in \epsfbox{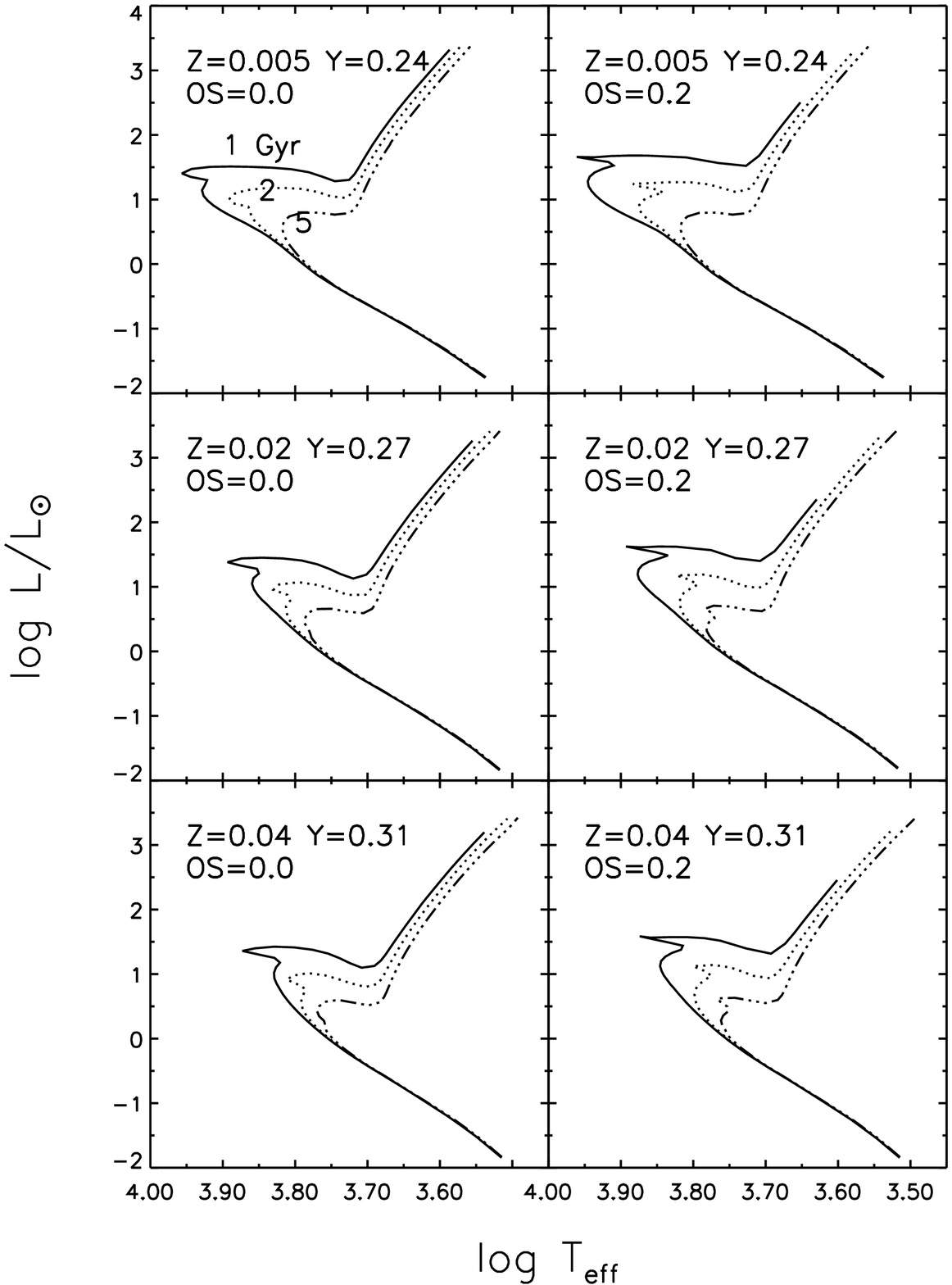}}
\centerline{\parbox{3.0in}{\small {\sc Fig. 3}
Part of the new Yale Isochrones that are used in this study.
Each panel shows 1, 2, and 5 Gyr isochrones with and without overshoot.
These isochrones are available from this paper through the Astrophysical
Journal.
}}
\vspace{0.2in}
\addtocounter{figure}{3}

	The post-RGB phase in the population synthesis, including the
horizontal branch (HB), has been constructed following the description of 
Yi et al. (1997b).
	The adopted input parameters are $\eta = 0.7$ (the mass loss efficiency
parameter in Reimers' empirical formula) and $\sigma_{HB} = 0.04$ \Ms (the 
HB mass dispersion parameter).
	The justification for these choices is provided in Yi et al. (1999).
	However, details of the post-RGB prescriptions hardly affect the 
overall integrated spectrum at small ages, which is why it is particularly
feasible to estimate the ages of giant elliptical galaxies at $z$ = 1 -- 2. 
	All models in this paper are based on the instantaneous starburst
hypothesis. 

	Synthetic CMDs are then convolved with the Kurucz spectral library 
(Kurucz 1992) to generate integrated spectra.
	Figure 4 shows the integrated spectra of 1, 2 and 3 Gyr models for
solar compositions compared to the observed data of LBDS\,53W091.
	The rest-frame UV spectrum of LBDS\,53W091 is from 
Spinrad et al. (1997).

	The $R$, $J$, $H$, \& $K$ magnitudes, obtained from the same source, 
have been converted into relative fluxes following Equation 1.
\begin{eqnarray}
\frac{F_{X}}{F_{3150}}  &=& \frac{F_{R}}{F_{3150}}\, [\frac{F_{X}}{F_{R}}]_{Vega}\,  10^{0.4 (R-X)}\cr\cr
		&=& 0.414\, [\frac{F_{X}}{F_{R}}]_{Vega}\, 10^{0.4 (R-X)},
\end{eqnarray}
where $F_{X}$ is the flux from an object of magnitude $X$ 
($R=24.5 \pm 0.2$, $J=20.5 \pm 0.1$, $H=19.5 \pm 0.1$, $K=18.7 \pm 0.1$) and
$F_{3150}$ is the mean flux in the wavelength range 3000 -- 3300 \AA.
	Vega's flux ratios have been computed from the Vega spectrum provided
in the Kurucz library. 
	They are $F_{J}/F_{R}=0.147$, $F_{H}/F_{R}=0.052$, and
$F_{K}/F_{R}=0.019$.
	The resulting relative fluxes of LBDS\,53W091 are 
$F_{R}/F_{3150}=0.414$, $F_{J}/F_{3150}=2.423$, $F_{H}/F_{3150}=2.153$, 
and $F_{K}/F_{3150}=1.643$, with approximately 20\%, 10\%, 10\%, \& 
10\% observational uncertainties, respectively.
	Our relative fluxes in $R$, $J$, and $K$ agree with those in 
Bruzual \& Magris (1997a) within 1 -- 2\% difference, but our $H$ band 
relative flux is about 10\% lower than theirs.
	However, this does not make an appreciable difference in the age 
estimate.

	Whether we can simply combine these two sets of different data
(the UV spectrum and the photometric data) may be questionable because 
photometry covered a larger area of the galaxy than spectroscopy did 
(Spinrad et al. 1997).
	We will discuss this effect in \S 5.3.
	Despite that, both the UV spectrum and the photometric data
indicate small ages between 1 and 2 Gyr consistently.
	
\placefigure{fig4}
\parbox{3.0in}{\epsfxsize=3.0in \epsfbox{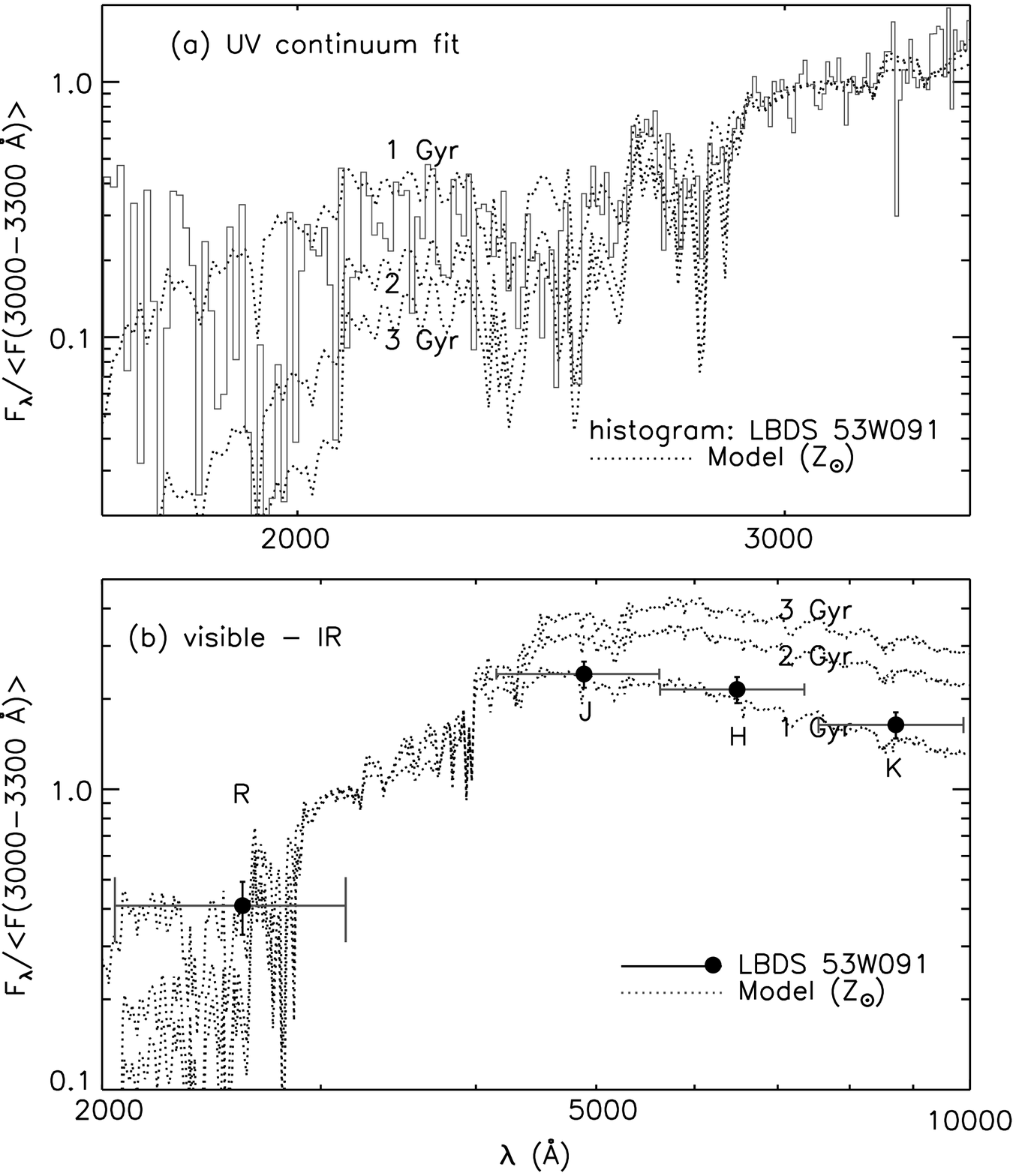}}
\centerline{\parbox{3.0in}{\small {\sc Fig. 4}
Model integrated spectra for solar abundance vs the observed data of 
LBDS\,53W091. The x-axis shows the rest-frame wavelength. All data of 
LBDS\,53W091 are from Spinrad et al. (1997). The four photometric data 
({\it filled circles}) have been derived from their $R$, $J$, $H$, \& $K$ 
magnitudes. The y-axis error bars are observational errors, and the x-axis 
ones show the effective band widths. The whole continuum is best matched if 
LBDS\,53W091 is 1 -- 2 Gyr old, when solar abundance is assumed.
}}
\vspace{0.2in}
\addtocounter{figure}{4}

\section{Effects of Convective Core Overshoot}

	Convective core overshoot (OS), the importance of which was first 
pointed out by Shaviv \& Salpeter (1973), is the inertia-induced penetrative 
motion of convective cells, reaching beyond the convective core as defined 
by the classic Schwarzschild (1906) criterion. 
	Stars develop convective cores if their masses are larger than
approximately 1.3 -- 1.5 \Ms, typical for the MSTO stars in 1 -- 5 Gyr-old 
populations, such as LBDS\,53W091.

	There has been a consensus for the presence of OS,
but its extent has been controversial.
	Thus, conventional stellar models often do not include OS.
	Since the advent of the OPAL opacities, a major improvement in the 
stellar astrophysics, various studies have suggested a modest
amount of OS; that is, OS $\approx 0.2$ \Hp, where \Hp is the pressure 
scale height (Stothers 1991; Demarque, Sarajedini, \& Guo 1994; 
Kozhurina-Platais et al. 1997).	

	OS has many effects on stellar evolution, but the most notable
ones are its effects on the shape of the MSTO and on the luminosity function.
	As shown in Figure 5, inclusion of a modest amount of OS causes a 
longer stretch of the MS before the blueward motion because it induces 
a larger supply of hydrogen fuel from the overshooting (``overmixing'') 
region into the core. 
	As a result, stars with OS ($= 0.2$ \Hp) stay longer in the core 
hydrogen burning stage, but in the MS rather than in the RGB.
	The table in Figure 5 lists the lifetime in the hydrogen burning phase
and in the RGB, both in Myr.

\placefigure{fig5}
\parbox{3.0in}{\epsfxsize=3.0in \epsfbox{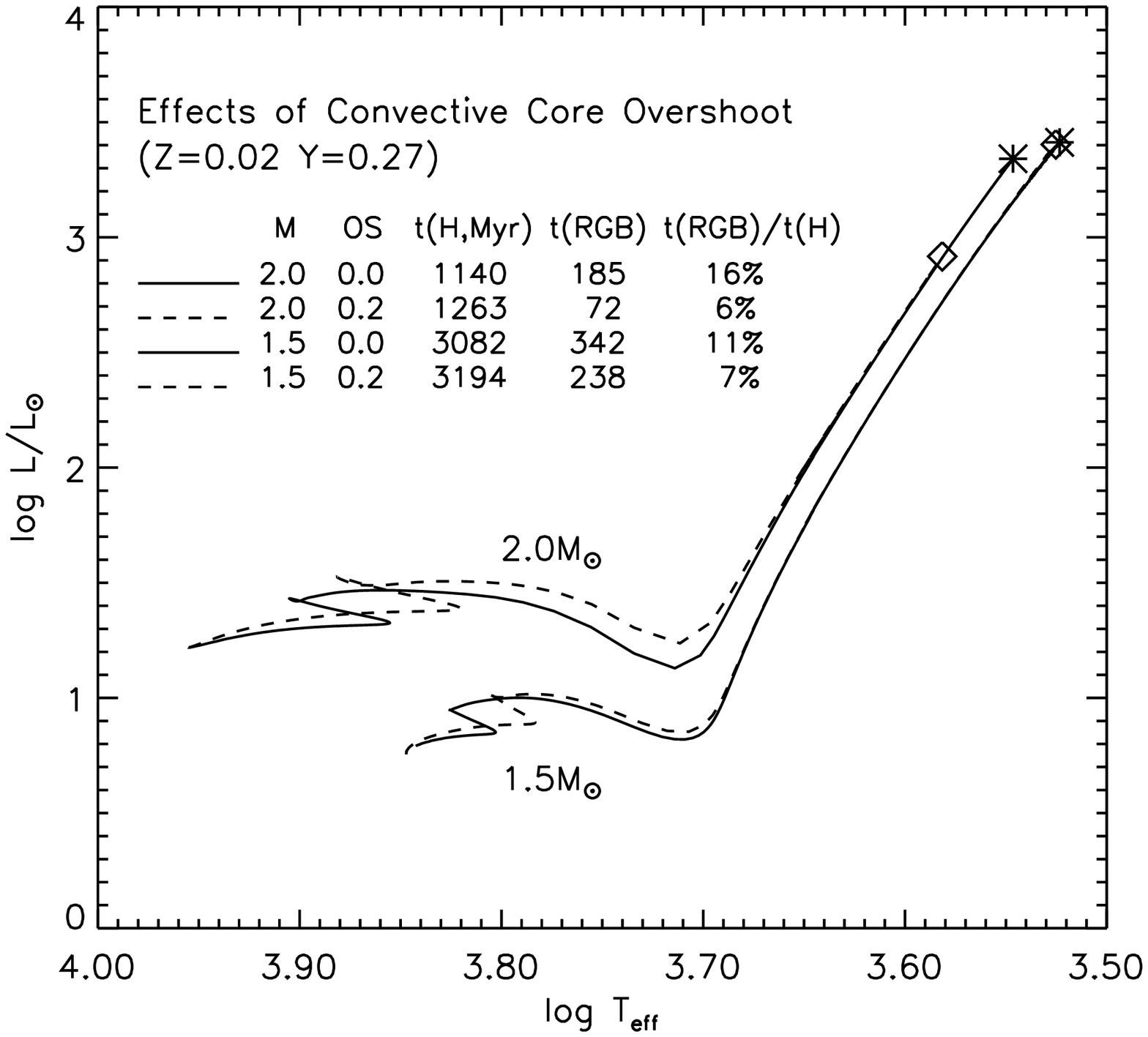}}
\centerline{\parbox{3.0in}{\small {\sc Fig. 5}
The effects of convective core overshoot (OS) to the stellar evolution
tracks of 1.5 and 2.0 \Ms. An increase in the adopted amount of OS causes
a longer lifetime in the hydrogen burning phase relative to that
in the helium burning phase and a longer lifetime in the MS than in the RGB.
The RGB tips are marked with $diamonds$ (with OS) and $asterisks$ (without OS).
}}
\vspace{0.2in}
\addtocounter{figure}{5}

\parbox{3.0in}{\epsfxsize=3.0in \epsfbox{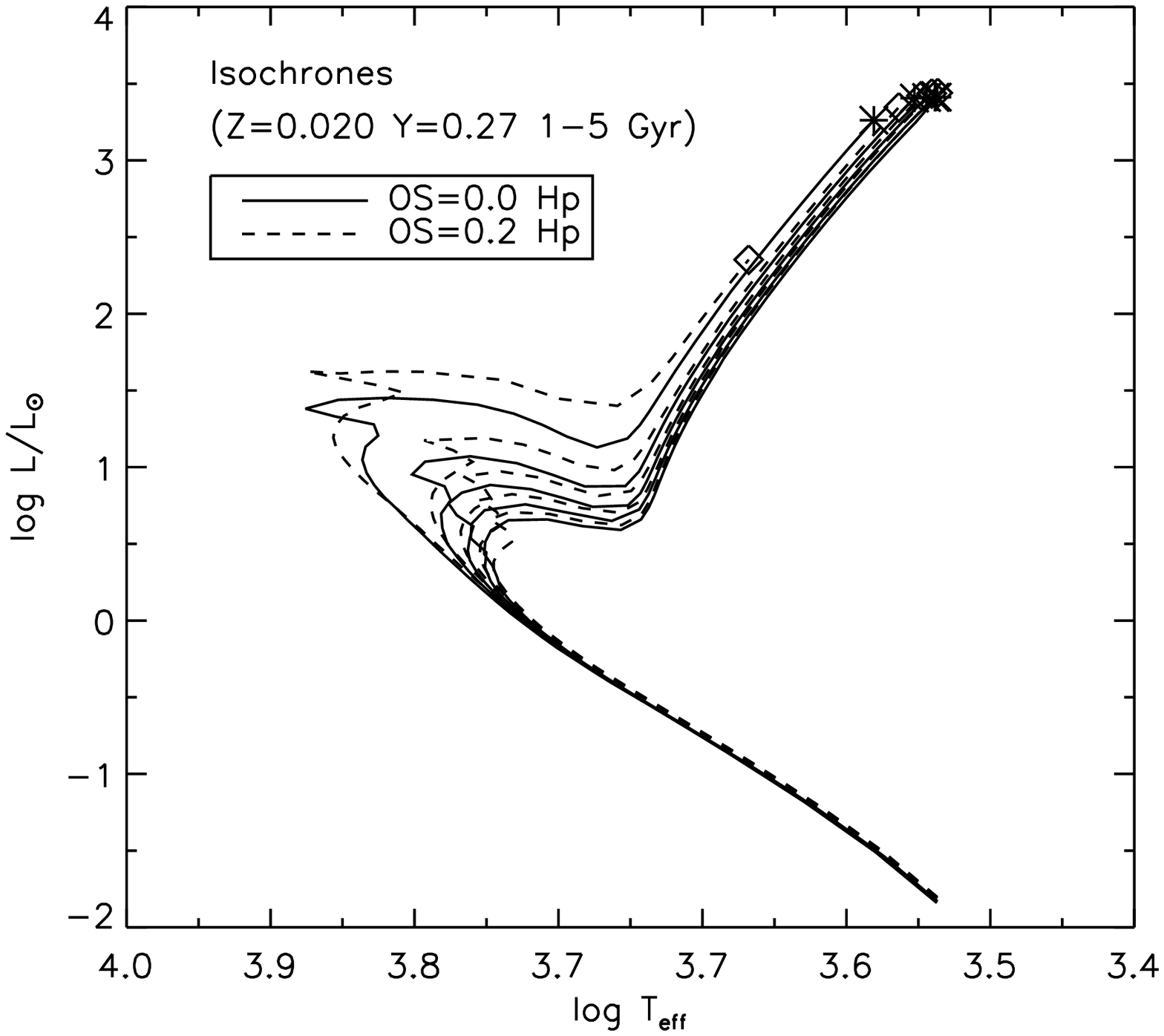}}
\centerline{\parbox{3.0in}{\small {\sc Fig. 6}
Isochrones with and without overshoot (OS).
The tip of the RGB has been marked with $diamonds$ (OS) and 
$asterisks$ (no-OS).
Note that the 1 Gyr isochrone with OS does not extend as far as its 
counterpart (without OS) does: the RGB tip in the 1 Gyr isochrone with OS 
is much fainter than that without OS. This causes a larger difference in 
the integrated spectrum, as shown in Figure 7.
}}
\vspace{0.2in}
\addtocounter{figure}{6}

	OS also causes stars to leave the RGB earlier.
	The {\em diamonds} in Figure 5 are the locations of the RGB tips
when OS is included, while {\em asterisks} are for the no-OS case.
	This directly affects isochrones, as shown in Figure 6.
	The RGB tips in the isochrones reach farther without OS.
	Because most of the visible flux comes from red giants, a decrease
in the RGB lifetime results in a lower visible flux.
	Figure 7 shows the integrated spectra with and without OS.
	One can see the impact of OS on the normalized integrated 
spectra of 1 Gyr models and thus on the age estimates.
	When the isochrones with OS are used in the population synthesis, 
the same observed integrated spectrum indicates a larger age, especially 
when the age is as small as 1 Gyr.

\placefigure{fig6}

\placefigure{fig7}
\parbox{3.0in}{\epsfxsize=3.0in \epsfbox{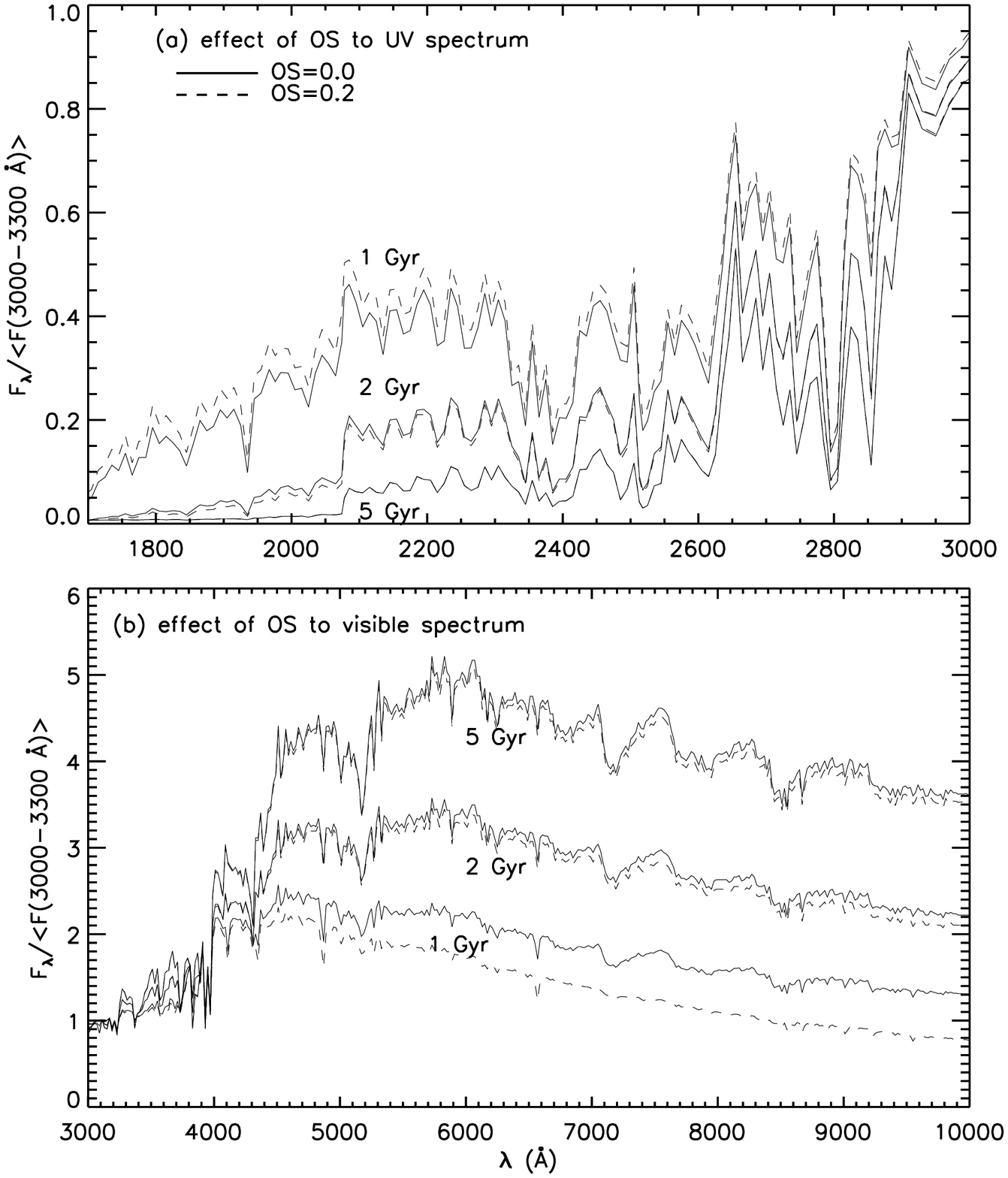}}
\centerline{\parbox{3.0in}{\small {\sc Fig. 7}
Model integrated spectra with and without overshoot (OS).
Note that the effect of OS is appreciable 
only in the young ($\lesssim 2$ Gyr) model's visible-IR spectrum.
}}
\vspace{0.2in}
\addtocounter{figure}{7}

\section{Effects of Metallicity Mixture}

	Most of the previous age estimates of LBDS\,53W091 were based on the 
single abundance population models, typically for solar composition.
	The solar abundance approximation has been popular for decades
because reliable spectral libraries were available only for solar compositions.
	The first obvious problem in this approximation is that there is 
little justification in the choice of solar composition in modeling giant 
elliptical galaxies.
	Observers have long believed that the majority of stars in 
giant elliptical galaxies are metal-rich (approximately twice solar) 
because of their extremely red colors and strong absorption lines.
	Moreover, most chemical evolution theories predict that giant 
elliptical galaxies would reach the current metallicity level within a few
tenths of a Gyr of the initial starburst (e.g., Kodama \& Arimoto 1997).
	This means that 1 -- 5 Gyr-old giant elliptical galaxies may
already have stars of various metallicities ranging $Z \approx$ 0 through 
perhaps 3 -- 4 times solar.
	Then, whether the single abundance approximation would be appropriate 
for modeling giant elliptical galaxies is questionable.

	Another problem is that the age estimate
based on continuum fits is quite sensitive to the metallicity adopted.
	For example, if we assume a high metallicity for LBDS\,53W091, 
the observed data of LBDS\,53W091 would indicate an age smaller than 1 Gyr 
(Figure 8-[c]).
	On the other hand, if the majority of stars in the galaxies at such
high redshifts were metal-poor, the same data would indicate a much larger
age by a fcator of 2 to 3 (Figure 8-[a]).
	This difference is caused by the opacity effects that increase
with increasing metallicity.
	As a result, the same observed spectrum can indicate significantly
different ages when the true metallicity is unknown.
	With such a large uncertainty in age estimate, it would be
extremely difficult to use the age estimates of distant galaxies to
constrain cosmology.

	The release of the improved Kurucz theoretical spectral library 
(Kurucz 1992) has finally enabled realistic modeling for compositions other 
than solar.
	Thus, it is our intention to construct more realistic models than
simplified solar abundance models.
	One can guess the effect of the use of metallicity mixtures
even with a cursory inspection of Figure 8.
	The fraction of metal-poor stars in conventional metallicity
mixture models is small.
	However, even with such small fractions of metal-poor stars,
a combination of coeval stars of various metallicities may have a substantial 
UV light contribution from metal-poor stars because of the opacity effects.

\parbox{3.0in}{\epsfxsize=3.0in \epsfbox{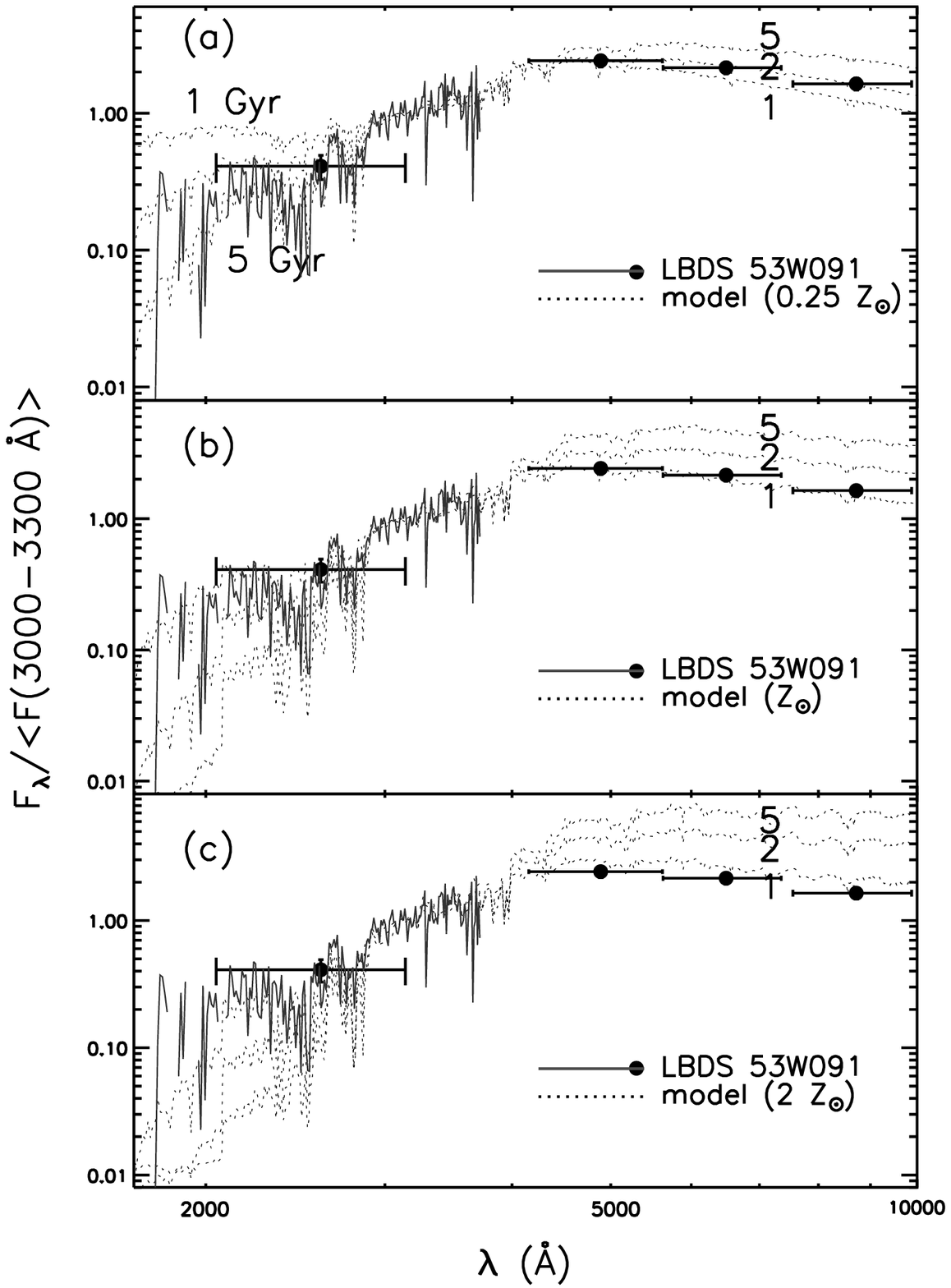}}
\centerline{\parbox{3.0in}{\small {\sc Fig. 8}
The effects of metallicity to the integrated spectrum.
Models are for 1, 2, and 5 Gyr and for no overshoot.
An error in the adopted metallicity leads to a large error in the age
estimate. One must adopt a realistic metallicity or metallicity
distribution.
}}
\vspace{0.2in}
\addtocounter{figure}{8}

	It should be noted that, while we try to make our models more 
realisitic than single abundance models by adopting metallicity mixtures and 
the Kurucz library, our models are still subject to the uncertainty in the 
spectral library.
	For example, Heap et al. (1998) noted that the Kurucz library does not
reproduce the correlations between some UV spectral strengths and 
optical colors found in the IUE sample.
	It is not our intention to claim that our age estimates based on 
composite models are free from such uncertainties in the spectral library.
	We only believe that the use of physically plausible
metallicity mixtures would result in the models superior to single abundance
models, other uncertainties remaining the same.

\placefigure{fig8}

	We have adopted the chemical evolution models of Kodama \& Arimoto 
(1997) that reasonably match the observed properties of present epoch giant 
elliptical galaxies.
	We have adopted both their metal-rich 
($<\!\!Z\!\!> \approx 0.04 \approx 2$ \Zs) ``simple'' and ``infall''
models for LBDS\,53W091.
	Their simple model is a closed-box model where the star formation time
scale is 0.1 Gyr.
	The galactic wind epoch, where the supernova-driven thermal energy 
exceeds the binding energy, is 0.353 Gyr from the beginning of the starburst.
	The infall model is based on the same configuration, except that
the unprocessed gas in the outskirt falls into the core with the time scale 
of 0.1 Gyr (see Kodama \& Arimoto 1997 for details).
	When only three metallicity bins ($Z = 0.005$, 0.02, 0.04)
are used, the fractions of metallicity groups are approximately
18\% for $Z = 0.005$, 24\% for $Z = 0.02$, 58\% for $Z = 0.04$ in the 
``simple'' model, and 12\% for $Z = 0.005$, 31\% for $Z = 0.02$, 
57\% for $Z = 0.04$ in the ``infall'' model.
	If giant elliptical galaxies at $z = 1$ -- 2 had significantly 
different metallicity distributions from these models, our composite models 
should be used for illustration purpose only.

	Figure 9 illustrates the likely light contributions from various 
metallicity groups of stars when the galaxy is 1.5 Gyr old.
	As shown in Figure 9-(a), the composite-``infall'' 
model with OS matches the overall data of LBDS\,53W091 reasonably well at this 
age. 
	As we expected, metal-poor stars are more efficient UV sources 
than metal-rich stars, while the opposite is true in the longer 
wavelength regions (Figure 9-[b]).
	The level of the light contribution from metal-poor stars is even 
higher when the ``simple'' (instead of ``infall'') model is used. 
	This is because ``simple'' models generally predict a 
larger fraction of metal-poor stars than ``infall'' models do.

\placefigure{fig9}
\parbox{3.0in}{\epsfxsize=3.0in \epsfbox{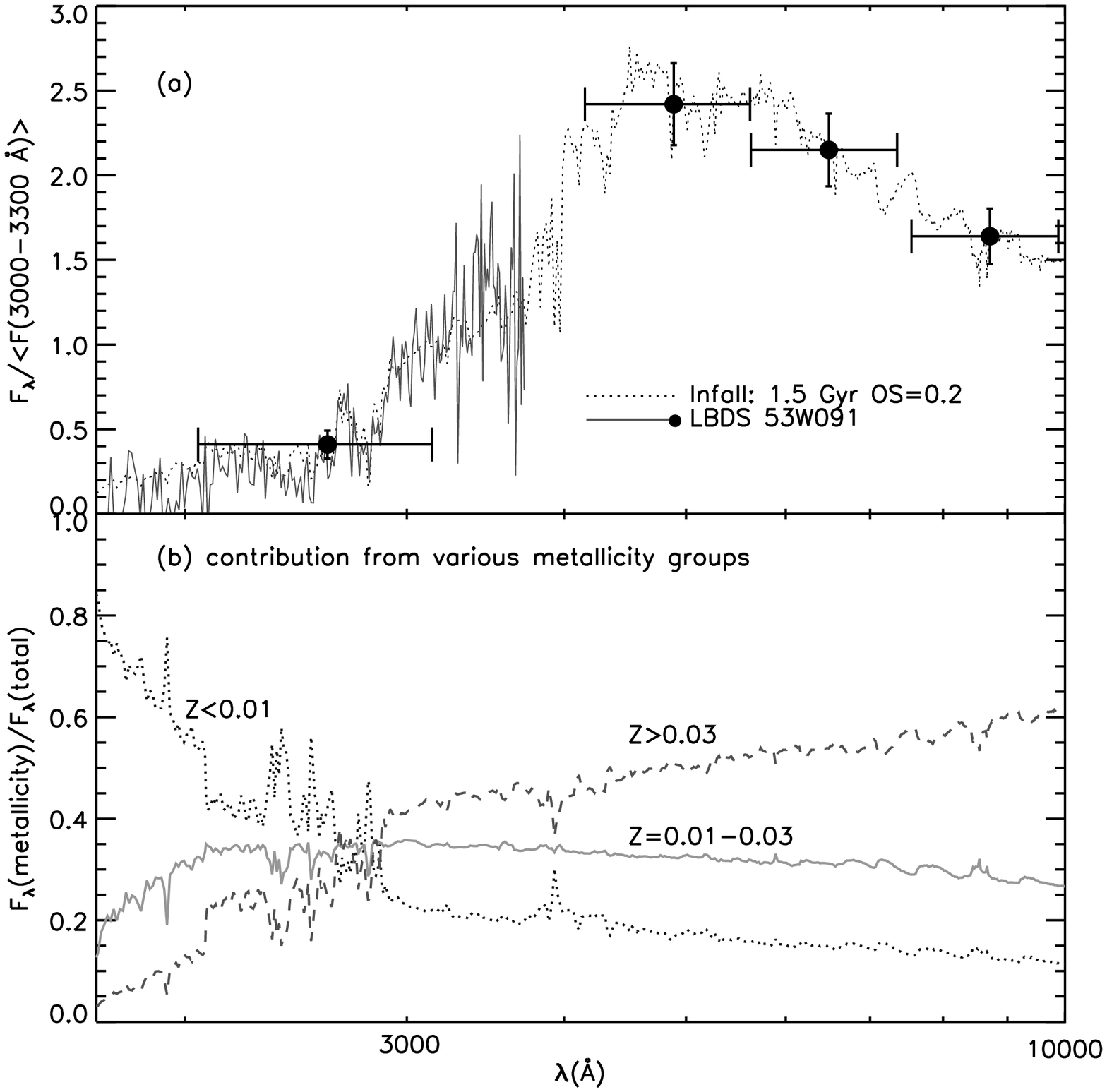}}
\centerline{\parbox{3.0in}{\small {\sc Fig. 9}
$top$: A composite-infall model of 1.5 Gyr of age fits the overall 
continuum of LBDS
53W091 reasonably well. $bottom$: While the visible spectrum is dominated
by metal-rich stars, the UV spectrum is strongly affected by metal-poor 
stars. Therefore, it is important to use a realistic metallicity distribution
in the galaxy population synthesis.
}}
\vspace{0.2in}
\addtocounter{figure}{9}

\section{Results}

	We perform $\chi^{2}$ minimization fits using the UV spectrum and
the photometric data of LBDS\,53W091, and the results are following.

\subsection{Fit to the UV Spectrum}

	We have carried out a weighted $\chi^{2}$ test
to the UV spectrum of LBDS\,53W091, using the following definition,
\begin{equation}
\chi^{2} = \frac{1}{N-2}\, \Sigma\, \frac{[f_{\lambda}(model) - f_{\lambda}(observed)]^{2}}{\sigma_{\lambda}^{2}},
\end{equation}
where $N$ and $\sigma_{\lambda}$ are the number of spectral bins 
and the observational errors, respectively.

	Figure 10 shows the measured values of reduced $\chi^{2}$ of 
various models.
	A smaller value of $\chi^{2}$ indicates a better fit.
	The spectra are normalized at 3150 \AA (the average flux in the range
of 3000 -- 3300 \AA), but the test was quite insensitive to the choice of the 
normalization point, as long as the normalization flux is the mean flux in a 
reasonably wide ($\Delta\lambda \gtrsim 50$ \AA) range.
	We modified the error value at 4973.47 \AA\, (1948.85 \AA\, in 
the rest-frame) with an approval from the observers. 
	It was 2 orders of magnitude smaller than other values, thus
$\chi^{2}$ tests were heavily dominated by the single point, producing
unreasonable results. 
	We assume that it was an artifact.
	We replaced it with an interpolated value from the adjacent bins.

\placefigure{fig10}
\parbox{3.0in}{\epsfxsize=3.0in \epsfbox{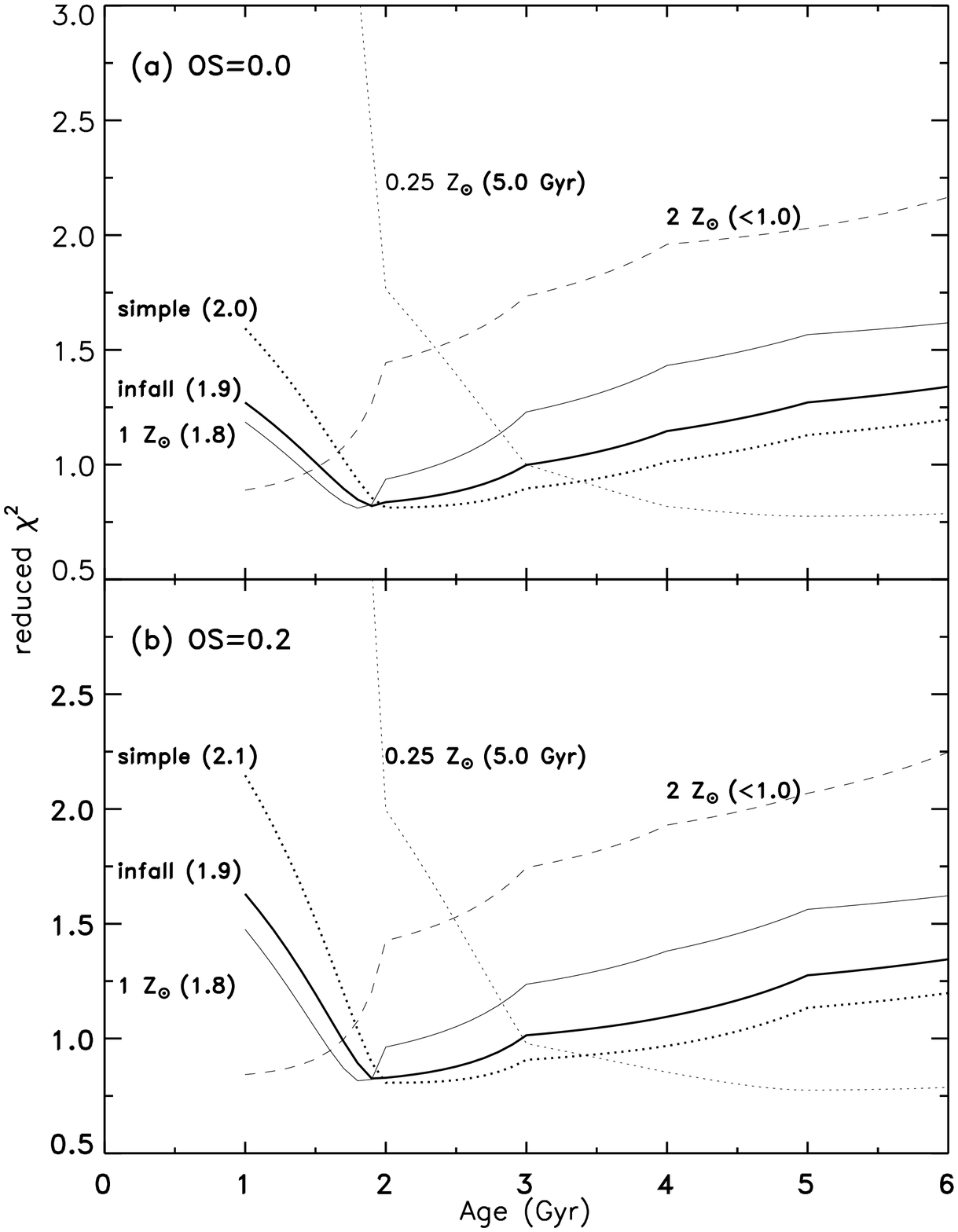}}
\centerline{\parbox{3.0in}{\small {\sc Fig. 10}
$\chi^{2}$ test of various models to the UV spectrum of LBDS\,53W091.
Each line is marked by the model name and the age of the model that shows the
minimum $\chi^{2}$.
Solar abundance models suggest 1.8 -- 1.9 Gyr and  composite models
suggest approximately 2 Gyr. Inclusion of overshoot has little impact.
}}
\vspace{0.2in}
\addtocounter{figure}{10}

	Figure 10 shows that OS has very little effect on the age estimates 
based on the UV spectrum, as illustrated earlier in Figure 7.
	The discussion of the effects of metallicity mixtures requires a bit 
of clarification first. 
	The mean metallicity of the majority of stars in giant elliptical 
galaxies is often believed to be approximately twice solar.
	If we use such a large metallicity for the single abundance 
population synthesis, then we would tend to underestimate the UV flux 
of this galaxy significantly and thus the age ($< 1.0$ Gyr).
	This is because even in such metal-rich ($<\!\!Z\!\!>\approx$~2~\Zs) galaxies 
the main UV sources are probably  metal-poor (see Figure 9).
	In this sense, the effect of the use of metallicity mixtures
is very large.
	As one can see in Figure 10, solar abundance models mimic
metal-rich (2 $Z_{\odot}$) composite models reasonably.
	Composite models indicate larger ages (1.9 -- 2.1 Gyr) than solar 
abundance models (1.8 Gyr) by approximately 10\%.
	This suggests that the solar abundance, used by various groups 
including Spinrad et al. (1997),  might be a reasonable approximation in 
the UV population synthesis of {\em young} giant elliptical galaxies,
even though it is still a factor of two smaller than the likely mean
metallicity of stars in giant elliptical galaxies (c.f., Angeletti \&
Gianonne 1999).
	When the three $\sigma$ confidence ranges are included,  
our age estimates based on the $\chi^{2}$ test using the UV spectrum alone
are approximately $1.8^{+0.2}_{-0.1}$ Gyr (solar abundance model with OS) 
through $1.9^{+0.6}_{-0.1}$ Gyr (infall model with OS), where the errors are
internal only and do not include observational errors.
	Thus, they are inconsistent with that of Spinrad et al. (1997) with 
more than a three $\sigma$ confidence (99.7\%).

\subsection{Fit to the Photometric Data}

	Photometric data are available for LBDS\,53W091 in $R$, $J$, $H$, 
\& $K$ magnitudes, covering approximately 2000 -- 9000 \AA\, in the rest-frame 
wavelength at the redshift of this galaxy ($z=1.552$).
	Photometric data may be less indicative of the age of a
population than the UV spectrum, but it is also less affected by the 
uncertainties in reddening and in metallicity.

	We have performed a weighted $\chi^{2}$ test on the four photometric
data points.
	The errors in magnitudes were also adopted from Spinrad et al. (1997).
	Figure 11 shows the reduced $\chi^{2}$ as a function of age.
	The models with minimum $\chi^{2}$ indicate ages of $< 2$~Gyr
unless the majority of stars in LBDS\,53W091 are very metal-poor.
	Composite models with OS = 0.2 suggest ages of approximately 1.5 Gyr.

	Unlike the analysis of the UV spectrum (\S 5.1), 
inclusion of OS raises the age estimates based on the photometric data 
substantially, 20 -- 50\%.
	This is because OS has a large impact on the visible -- IR flux 
when normalized to the UV (see Figure 7).
	If no OS is adopted, composite models suggest very small ages
($<$ 1.0 Gyr).
	The effect of inclusion of metallicity mixtures is also substantial.
	When the mean metallicity of stars in LBDS\,53W091 is assumed to be
twice solar, our models with OS suggest 25\% larger ages when
metallicity mixtures are included.
	This is because composite models contain some metal-poor stars
which match the $R-J$ color (i.e., rest-frame UV continuum) at older
ages than metal-rich stars do.

	When we use composite models with OS, our analysis on the photometric
data of LBDS\,53W091 suggests 1.5 $\pm$ 0.2 Gyr, 
which is significantly smaller than the age estimate
of Spinrad et al. but consistent with that of Bruzual \& Magris.
	This value is somewhat smaller than our estimate from the UV
analysis in the previous section.

\placefigure{fig11}

\subsection{The New Age estimate of LBDS\,53W091}

\placetable{tbl-2}
\begin{table*}
\caption{Age estimates (Gyr) of LBDS\,53W091 based on $\chi^{2}$ test\tablenotemark{a}.} \label{tbl-2}
\begin{center}
\begin{tabular}{cccc}
\tableline
\tableline
Metallicity & OS & estimate using UV spectrum & estimate using photometry \\
\tableline
0.25\Zs                 & 0.0 & $5.0^{+0.6}_{-0.3}$ & $2.1^{+0.4}_{-0.2}$ \\
0.25\Zs                 & 0.2 & $5.0^{+0.6}_{-0.2}$ & $2.5^{+0.3}_{-0.5}$ \\
\Zs                     & 0.0 & $1.8 \pm 0.1$     & $1.4^{+0.2}_{-0.3}$ \\
\Zs                     & 0.2 & $1.8 \pm 0.1$     & $1.7 \pm 0.1$ \\
2\Zs                    & 0.0 & $ \lesssim 1.0$   & $ \lesssim 1.0$ \\
2\Zs                    & 0.2 & $ \lesssim 1.0$   & $1.1^{+0.2}_{-0.1}$ \\
simple\tablenotemark{b} & 0.0 & $2.0^{+0.2}_{-0.1}$ & $ \lesssim 1.0$ \\
simple\tablenotemark{b} & 0.2 & $2.1^{+0.3}_{-0.2}$ & $1.5^{+0.2}_{-0.1}$ \\
infall\tablenotemark{b} & 0.0 & $1.9 \pm 0.1$ & $ \lesssim 1.0$ \\
infall\tablenotemark{b} & 0.2 & $1.9^{+0.2}_{-0.1}$ & $1.5 \pm 0.1$ \\
\tableline
\end{tabular}
\end{center}
\tablenotetext{a}{Errors are systematic errors for one $\sigma$. 
The resolution of our models in age is 0.1 Gyr.}
\tablenotetext{b}{Composite models from Kodama \& Arimoto (1997): $<\!\!Z\!\!> \
approx 2 Z_{\odot}$}
\end{table*}

	Table 2 shows the list of our age estimates based on the two data sets.
	They range approximately 1 -- 2 Gyr unless extreme metallicities 
are assumed.
	Figure 12 shows a sample fit.
	The observed data are shown as a continuous line and filled circles.
	The overplotted models are those with the infall mixture and OS, 
which are probably more realistic than single abundance models with no OS.
	In the UV, only the model with the minimum $\chi^{2}$ (1.9 Gyr) 
is shown for clarity.
	In the longer wavelength regions, multiple models with various ages 
are plotted over the photometric data.
	This figure demonstrates the difference between the age estimates
from the analyses of the UV spectrum and of the photometric data.

\parbox{3.0in}{\epsfxsize=3.0in \epsfbox{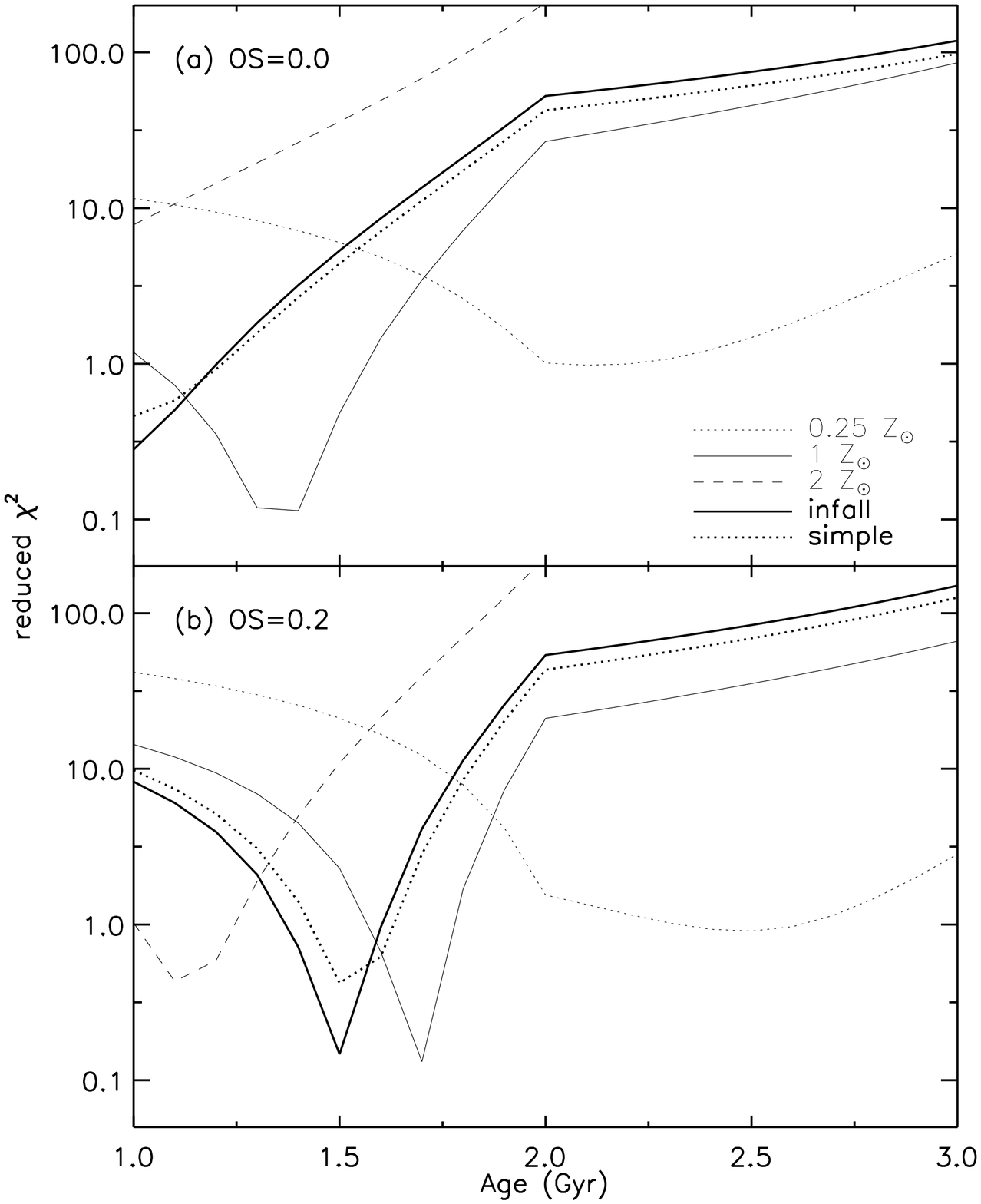}}
\centerline{\parbox{3.0in}{\small {\sc Fig. 11}
$\chi^{2}$ test of various models to the photometric data of LBDS\,53W091. 
Composite models with OS, the most realistic models, reach their minimum
$\chi^{2}$ near the age of 1.5 Gyr.
}}
\vspace{0.2in}
\addtocounter{figure}{11}

\placefigure{fig12}
\parbox{3.0in}{\epsfxsize=3.0in \epsfbox{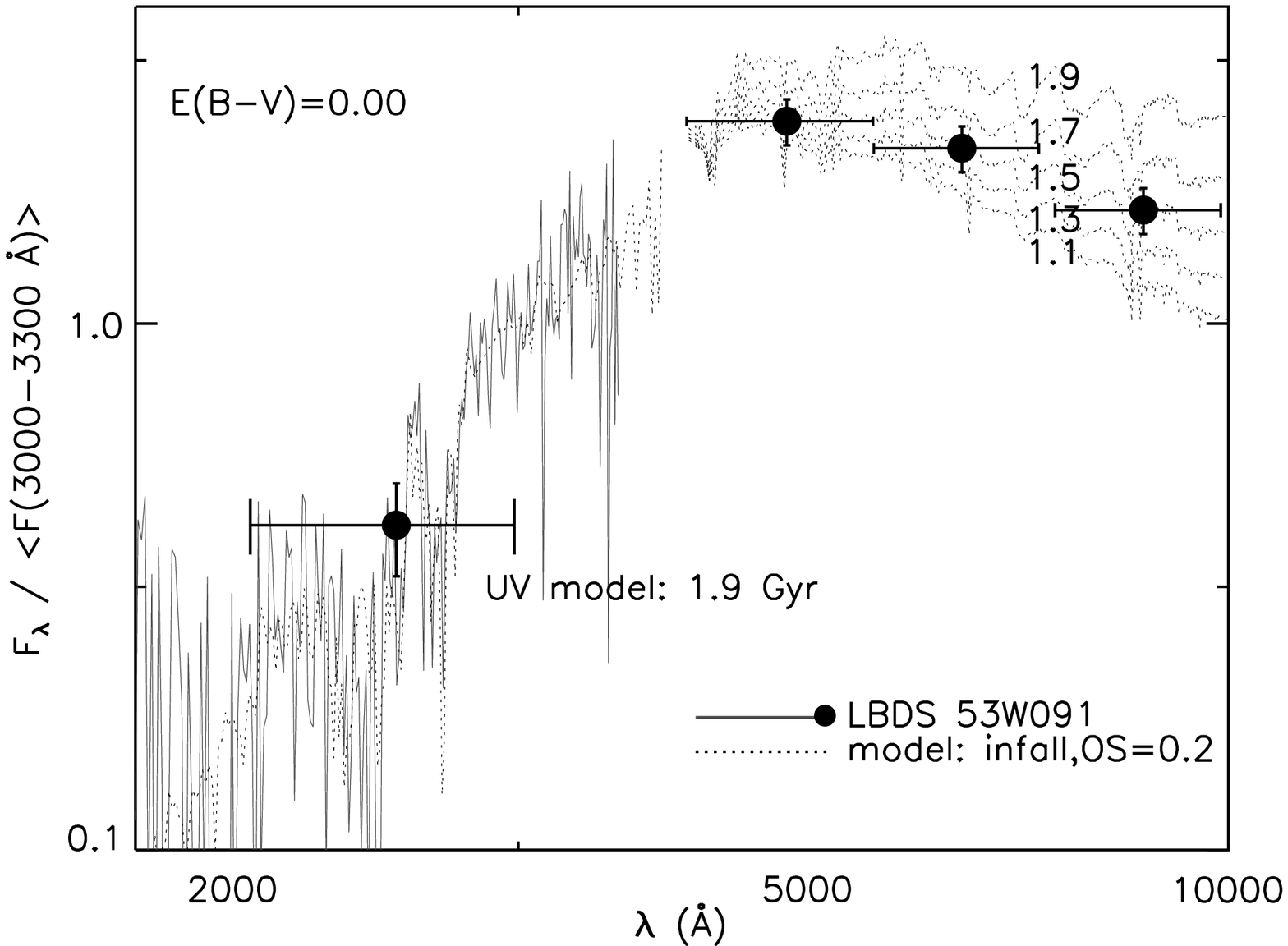}}
\centerline{\parbox{3.0in}{\small {\sc Fig. 12}
The model spectra compared to the observed data of LBDS\,53W091.
In the UV, only the best matching model (with the minimum $\chi^{2}$)
is overplotted, while in the longer wavelength range multiple models
with various ages (shown in Gyr on lines) are shown.
The UV spectrum indicates somewhat larger age than the the photometric data
in the longer wavelength regions. This may be due to radial gradients in 
age and/or metallicity, or to reddening.
}}
\vspace{0.2in}
\addtocounter{figure}{12}

	This difference between the two age estimates may appear substantial
to some readers.
	However, such a difference would not be unnatural if the data have 
been affected by some reddening, whether it is Galactic or internal.
	If we use the extinction curve of Cardelli, Clayton, \& Mathis
(1989), in conjunction with that of O'Donnell (1994), a moderate amount of 
Galactic reddening, $E(B-V) = 0.04$, reduces the difference between 
these two age estimates.
	In fact, the Galactic foreground 
extinction is estimated at 0.03 mag by Schlegel, Finkbeiner, \& Davis (1998),
along this line of sight ($RA$ = 17:21:17.84, $Dec$ = 50:08:47.7, 1950.00; 
$l$ = 76.8684, $b$ = 34.4531).
	Then, both the UV spectrum and the photometric data become consistent
with an age of 1.3 Gyr (Figure 13).

	It is also probable that some of this difference 
comes from the fact that the light sources of the UV spectrum are older 
and/or more metal-rich than those of the photometric data, because the UV 
spectroscopy covered a smaller area near the center of this galaxy 
(Spinrad et al. 1997).
	Then, composite models with radial gradients in age and metallicity 
would be more realistic to explain the observed data. 
	
	Our UV-based age estimates are in principle lower limits because
the UV spectrum is likely to be dominated by the youngest population.
	However, if the star formation time scale was only an order of
0.1 Gyr, as many galactic chemical evolution models for
giant elliptical galaxies suggest (e.g., Kodama \& Arimoto 1997),
this age spread effect may not be large.
	It is important to note that our UV-based age estimates are 
larger than those from the photometric data.
	Firstly, this may indicate the presence of at least some reddening.
	We have shown above that a small amount of reddening reconciles
the difference between the two age estimates.
	Secondly, if there is any substantial age spread in LBDS 53W091,
the UV-based age estimates should be smaller than the visible-based ones,
because the shorter wavelength spectrum is more dominated by the younger stars.
        For example, if a galaxy experiences a starburst that lasts for 
1 Gyr at a constant rate centered at 2 Gyr before the observation, 
its integrated UV spectrum would be matched best by a single burst model
of an age of 1.8 -- 1.9 Gyr instead of 2.0 Gyr. 
        In some {\em ad hoc} star formation scenarios, the age underestimation
can of course be larger.
        However, in the case of LBDS\,53W091, UV-based age estimates
are generally larger than the visible-based ones, which is opposite to the 
expectation from the population with any age spread.
        Thus, this likely implies {\em very little age spread}
in LBDS\,53W091 and {\em the presence of some reddening}, 
unless it is entirely due to the aperture effect. 
	Then, our UV-based age estimates may even be an upper limit,
rather than a lower limit.
	Without knowing the accurate amount of age spread and reddening
and the level of accuracy of the models, it is not clear whether our 
estimates can constrain lower limits or upper limits.  
	
\placefigure{fig13}
\parbox{3.0in}{\epsfxsize=3.0in \epsfbox{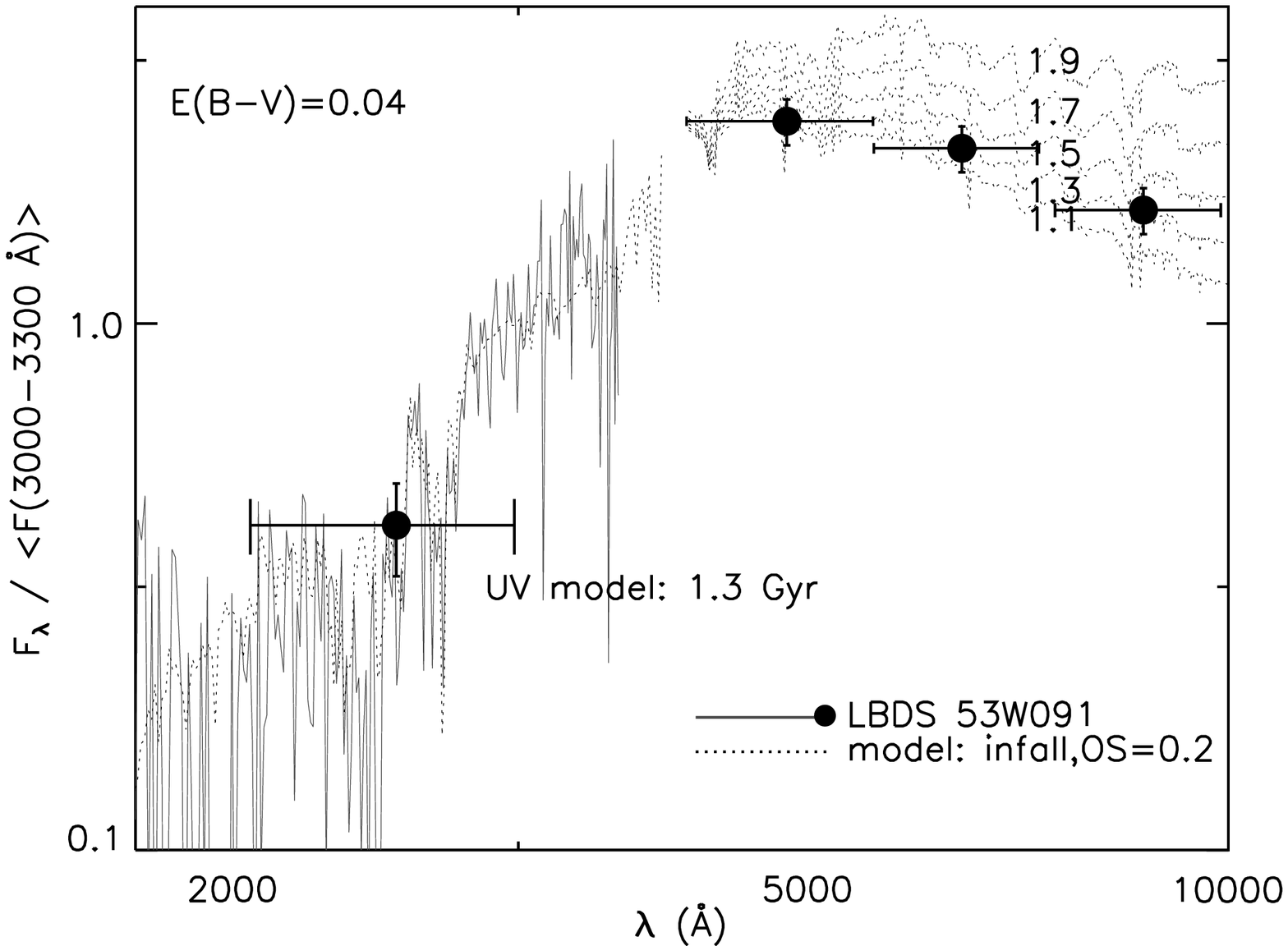}}
\centerline{\parbox{3.0in}{\small {\sc Fig. 13}
Same as Figure 12, but with the effect of reddening included.
The models are reddened according to the extinction curves of Cardelli et al.
(1989) and of O'Donnell (1994).
A 1.3 Gyr model matches both UV and visible -- IR data reasonably.
}}
\vspace{0.2in}
\addtocounter{figure}{13}

\section{Origin of the Disagreement among Various Models}

        The large difference in age estimate from Spinrad et al. and from
this study is mainly due to the significant difference in the model 
integrated spectrum.
        Figure 14 shows the comparison between the latest Jimenez 
models\footnote{downloaded from Jimenez' ftp site in May, 1999}
(the preferred models in the analysis of Spinrad et al.)
and the Yi models, both for the solar composition with no OS.
	The Jimenez models shown here are the new ones based on the 
stellar evolutionary tracks with finer 
mass grids and so probably improved over those used in Spinrad et al. (1997).
	At the time of this study, only his new models were made
available to us.

\placefigure{fig14}
\parbox{3.0in}{\epsfxsize=3.0in \epsfbox{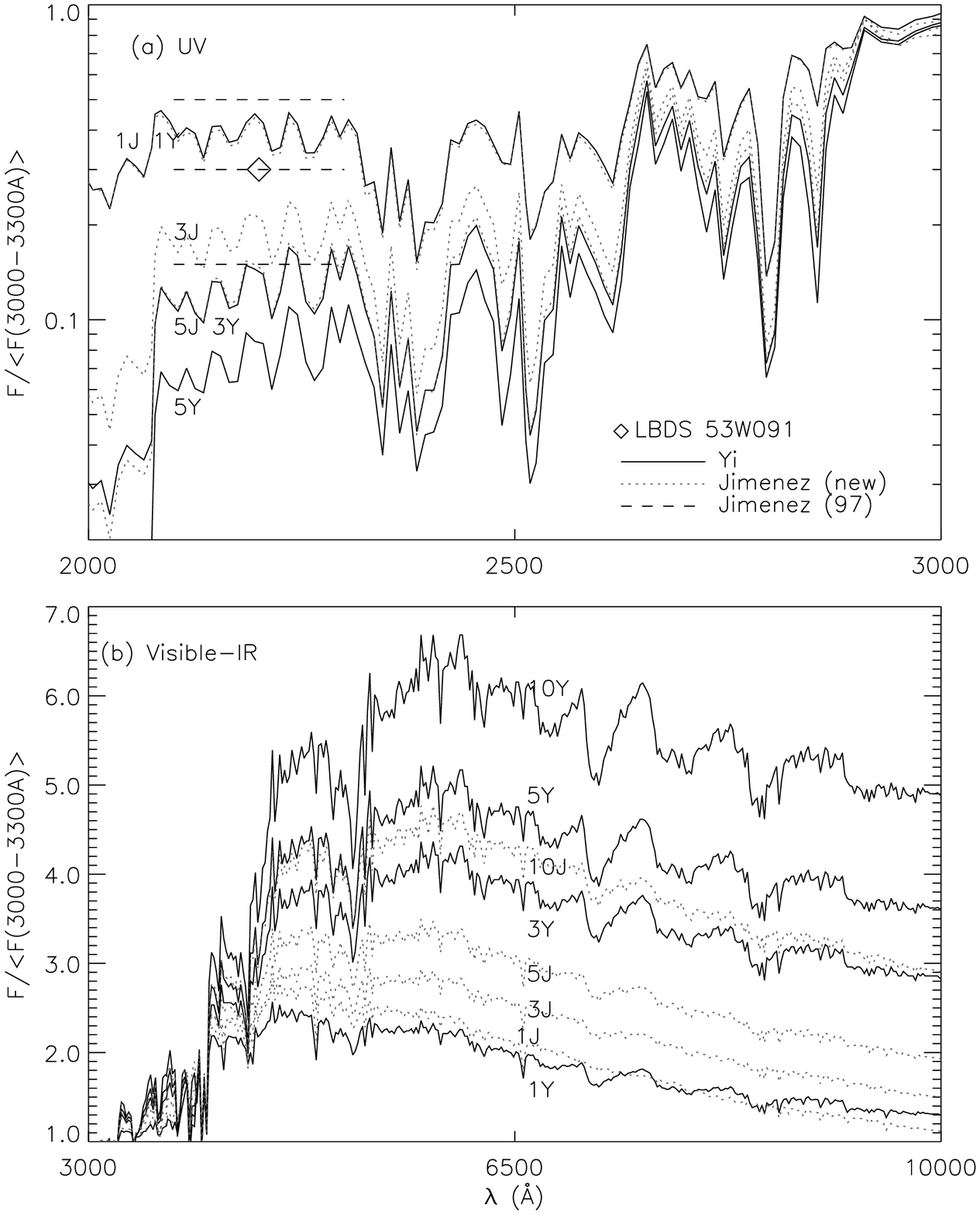}}
\centerline{\parbox{3.0in}{\small {\sc Fig. 14}
Comparison between the Yi models (marked ``Y'') and the Jimenez models 
(marked ``J'') for solar composition. Ages in Gyr are marked on each model.
Both in the UV (a) and in the visible -- IR (b), the Jimenez models are
substantially bluer. 
}}
\vspace{0.2in}
\addtocounter{figure}{14}

        In the UV, Jimenez' 1 Gyr model (1J) agrees with Yi's model (1Y) 
remarkably well.
	However, they substantially disagree for larger ages:
Yi's 2, 3, 5 Gyr models are close to Jimenez' 3, 5, 10 Gyr models
(not all models shown here for clarity).
	In fact, Jimenez' 3 Gyr and 5 Gyr models are nearly on top of
Yi's 2 Gyr and 3 Gyr models.
	One can easily understand why Spinrad et al. and we are achieving
such different age estimates.
	An open diamond in Figure 14-(a) is an approximate relative flux
of LBDS 53W091 normalized at 3150 \AA.
	This relative flux of LBDS 53W091 is closely reproduced by the 1.4 Gyr
model when the Yi models are used (see also Table 2) or the 
$1.9^{+0.2}_{-0.1}$ Gyr models when the new  Jimenez models are used 
(based on the reduced-$\chi^{2}$ tests shown in Equation 2).

	Also plotted in Figure 14-(a) are the relative fluxes from the 
earlier version Jimenez models (horizontal dashed lines) 
used in Spinrad et al. (1997).
	Because we did not have access to this version Jimenez 
models, we read off the relative fluxes from the Figure 14 of Spinrad et al.
(1997).
	The three dashed lines are from the top his 1, 3, and 5 Gyr models,
respectively.
	Note that his early version models are substantially bluer than
his new version.
	One can see that the early version Jimenez models match the
LBDS 53W091 data approximately at 3 Gyr, as Spinrad et al. (1997) suggested.
	The fit to the UV spectrum using the new Jimenez models,
which suggests an age estimate of 1.9 Gyr, is shown in Figure 16-(a).
	Readers are encouraged to compare Figure 16-(a) to 
Figure 14 of Spinrad et al. (1997).
	As shown in Figure 15, the Yi models are somewhat redder than 
the 1999 version Bruzual \& Charlot (BC, private communication) models, 
probably because the BC models assume a larger UV light contribution from 
post-asymptotic giant branch (PAGB) stars, but the overall agreement is good.
	If the new Jimenez models are improved over his previous version
used by Spinrad et al., all three groups now suggest rather consistent
age estimates, between 1 and 2 Gyr.

\placefigure{fig15}
\parbox{3.0in}{\epsfxsize=3.0in \epsfbox{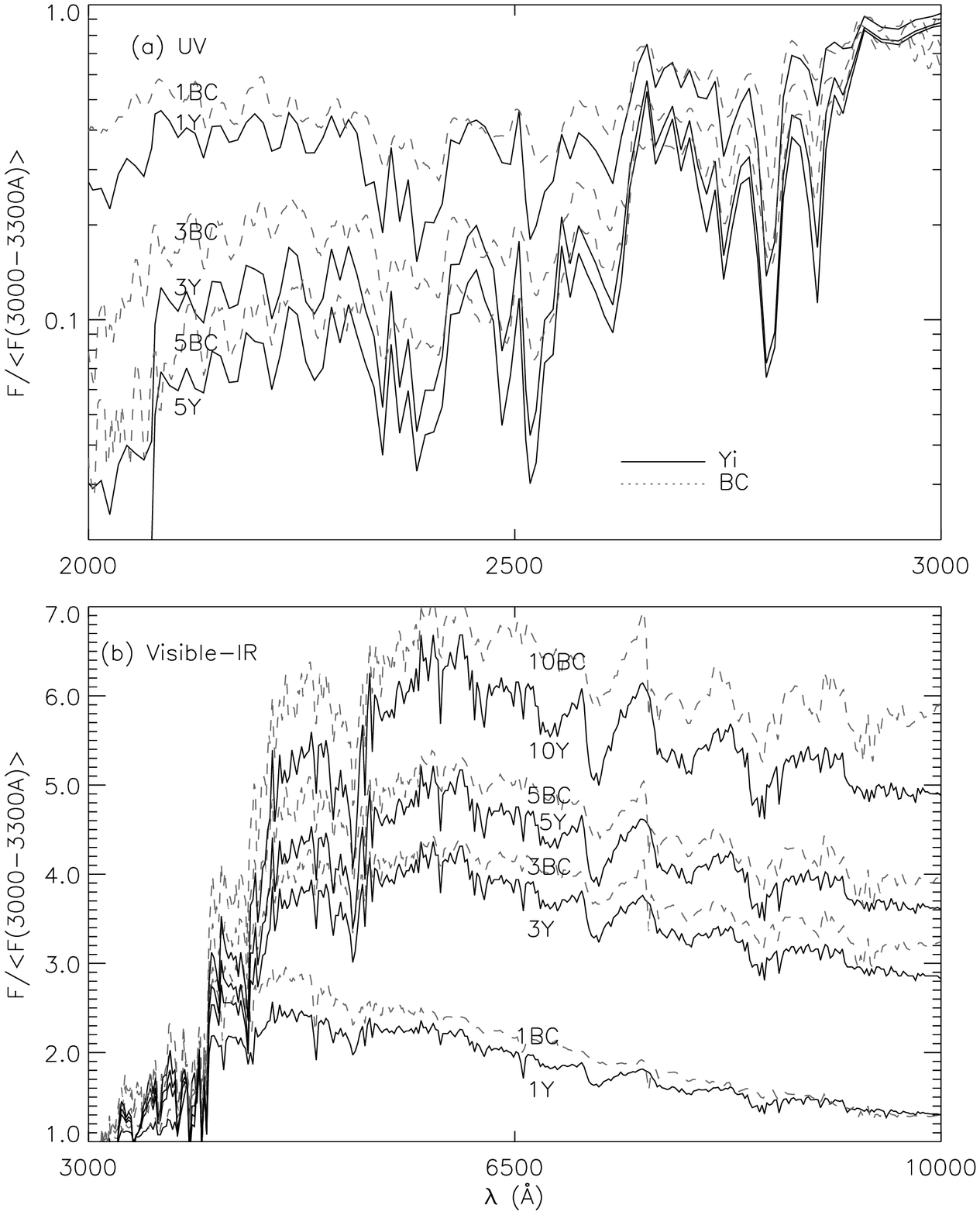}}
\centerline{\parbox{3.0in}{\small {\sc Fig. 15}
Comparison between the Yi models (marked ``Y'') and the 1999 version 
Bruzual \& Charlot models (marked ``BC'') for solar composition.
They are in good agreement, except that the BC models are somewhat redder
because they are based on the corrected Kurucz spectral library (see text).
}}
\vspace{0.2in}
\addtocounter{figure}{15}
	
        In the visible -- IR, the difference between the new Jimenez 
models and the Yi models is even larger.
        Figure 14-(b) compares their 1, 3, 5, and 10 Gyr models. 
        Jimenez' 1, 3 and 5 Gyr models are very close to one another, while
Yi's models are quite well separated.
        Jimenez' 5 and 10 Gyr models are quite close to Yi's 2 and 4 Gyr 
models, respectively.
	The Yi models are in close agreement with the BC models, as shown
in Figure 15-(b).
	The BC models are based on a modified Kurucz spectral library
(Lejeune, Cuisinier, \& Buser 1997) and thus somewhat redder than the Yi models.
	The difference between the Jimenez models and the Yi models, however,
is not the only cause for Spinrad et al. to get such a large age estimate, 
i.e., 2.5 Gyr, from their analysis on the photometric data.
	It was also caused by the fact that they used only $R-K$, omitting
$J$ and $H$ magnitudes in their analysis.
	Figure 16-(b) shows that even the Jimenez models would have suggested
a substantially smaller age if the whole photometric data had been used in 
their analysis. 
	When we use the new Jimenez models and the reduced-$\chi^{2}$ test
on the all of the four photometric data points, the best model indicates
$1.9^{+0.5}_{-0.7}$ Gyr, which is in good agreement with the estimate from
the UV analysis.
	Then again, all three groups scrutinized here are suggesting
consistent age estimates of 1 -- 2 Gyr.
	Such an agreement, at least on the age of LBDS 53W091, is possible 
because, for small ages ($\lesssim 2$ Gyr), the new Jimenez models differ 
from the Yi and the BC models only slightly.

\placefigure{fig16}
\parbox{3.0in}{\epsfxsize=3.0in \epsfbox{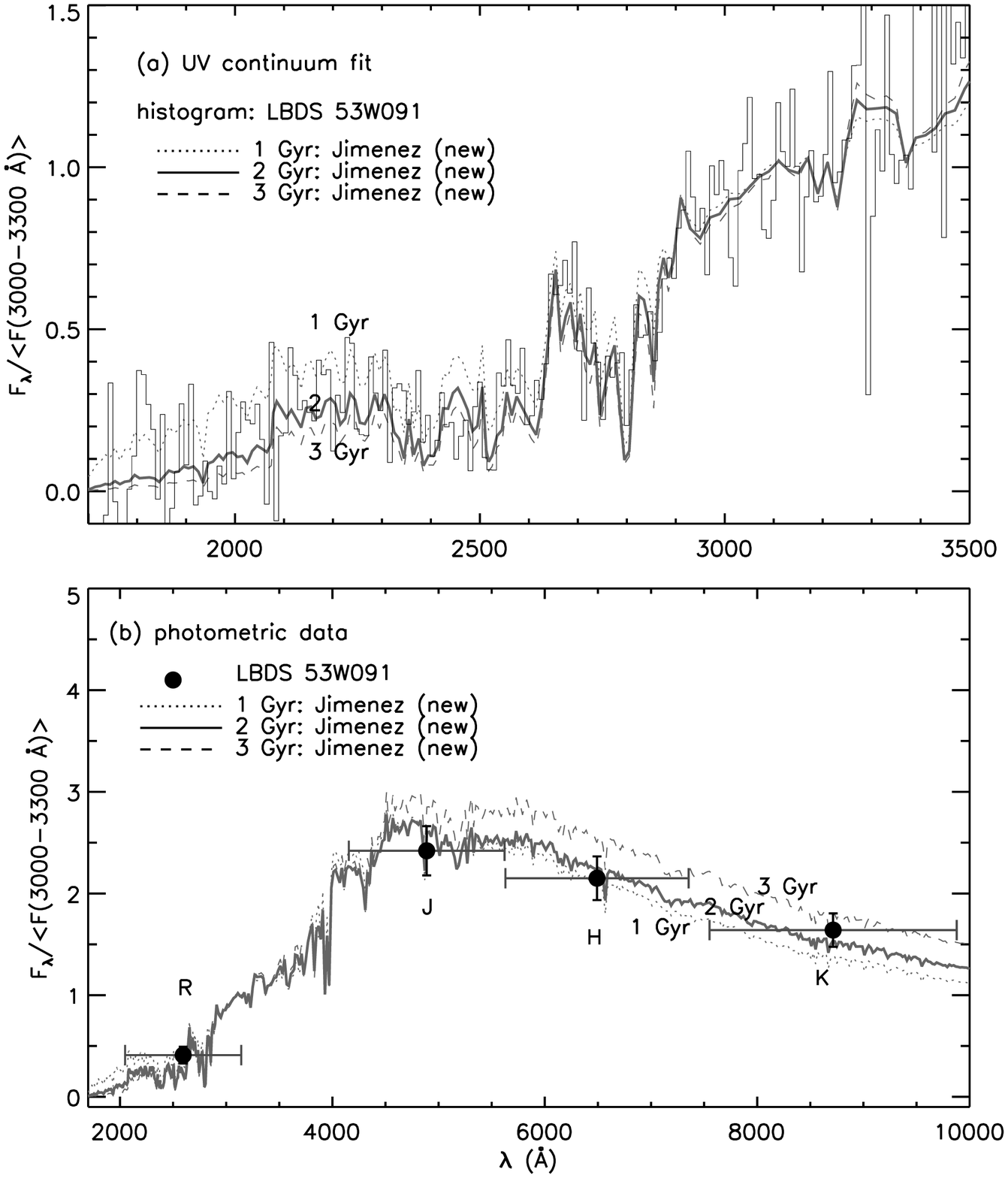}}
\centerline{\parbox{3.0in}{\small {\sc Fig. 16}
The fit to the data of LBDS 53W091 using the new Jimenez models.
Both its UV spectrum and the photometric data consistently suggest
approximately 1.9 Gyr.
}}
\vspace{0.2in}
\addtocounter{figure}{16}

	One may then say that the age discrepancy on LBDS 53W091 has 
been resolved. 
	Despite this apparent resolution, it is still quite disturbing to know
that there are significant disagreements between the Jimenez models and
the Yi (and the BC) models at larger ages.
        If such a large disagreement is due to the uncertainties in the 
input physics, we would have to admit that we are not ready to estimate ages 
of the bulk of stellar populations via continuum fitting.

        A notable disagreement can easily be introduced by the details in 
the population synthesis technique.  
        For example, inaccurate interpolations between tracks, whether they 
take place during the isochrone construction or directly in the population 
synthesis, may cause significant difference especially in the visible -- IR 
spectrum.
        The large impact of inaccurate mass interpolation has already been
pointed out in \S 2 and perhaps demonstrated by the change in the
two versions of the Jimenez models.

        Despite such complexities in modeling the visible -- IR flux, 
it is not true that model spectra in the visible -- IR are generally unreliable
(see Dunlop 1998 for a different opinion).
        For example, the age estimate using the old (1987) Revised Yale
Isochrones, where RGB tips were computed less accurately than in the
current Yale Isochrones used in this study, is different from our current 
estimate only by 20\%.
	The difference between the estimates of Bruzual \& Magris (1997a),
based on the BC models, and ours is also small, even though the two groups 
use different stellar evolutionary tracks.
	Besides all these, the conventional population synthesis models
beautifully match the overall spectra (near-UV through near-IR) of globular 
clusters (e.g., Bruzual et al. 1997) at their accepted ages.
        In short, the visible -- IR flux is more sensitive to the 
uncertainties in the stellar models and in the population synthesis 
(including isochrone construction), but the level of uncertainty
should be less than 20\% or so in age.
        Thus, it is still unclear where such a large difference in the model 
spectrum as present between the Jimenez models and the Yi (and the BC) 
models comes from.
        It is unlikely to be caused by the difference in the input physics in 
the stellar models, because the stellar evolution theory is already quite 
well established.
        All these three groups use the Kurucz spectral library or
some hybrid versions originated from it, and, thus, it is also unlikely that
much of the difference can be attributed to the spectral library.
        The source of the discrepancy will be known only when these
models are compared step by step with each other.

        While it may be extremely difficult to directly compare different 
population models and find a more realistic model, it should be possible 
to test models by matching the observational properties of the objects whose 
ages are reasonably well known.
	Good examples include the sun, M32, and Galactic globular clusters.
	We are presenting the results of the tests on the sun and M32 only, 
because globular clusters have already been modeled successfully by a 
number of studies including that of Bruzual et al. (1997).

	The following test results may not appear relevant to some readers,
because the sun and M32 are likely significantly older than LBDS 53W091. 
	However, we would like to provide some useful 
sample tests that can easily be used to validate population synthesis models.

\subsection{Test on the sun}

	The age of the sun, approximately $4.53 \pm 0.4$ Gyr 
(see Guenther \& Demarque 1997 for references), is one of the best 
constrained quantities in astronomy.
	In a 5 Gyr coeval population, most of the UV light is still 
produced by MS stars.
	Thus, a 5 Gyr-old solar abundance population should exhibit a UV flux 
similar to that of the sun.
	Figure 17 shows the fits to the theoretical solar spectrum 
(2000 -- 3500 \AA) from the Kurucz library using the Yi models (a) and the 
new Jimenez models (b).
	Our test on the sun is not weighted with observational 
errors because the solar spectrum is a theoretical one.
	In this sense, we are measuring merely the mean square difference 
($MSD$) between the model and the data instead of $\chi^{2}$.
	The formula for the $MSD$ test is shown in Equation 3.
\begin{equation}
MSD = \frac{1}{N-2}\, \Sigma\, \frac{[f_{\lambda}(model) - f_{\lambda}(object)]^{2}}{f_{\lambda}(object)},
\end{equation}
where $N$ is the number of spectral bins.

	The best fitting model is a $5.0 \pm 0.1$ Gyr model when the Yi 
models are used.
        This slight disagreement in the age estimate from the generally 
accepted solar age is perfectly expected, because we are matching a single 
stellar spectrum with those of composite (in the sense of containing 
MS, red giants, and $etc.$) stellar populations and the light contribution
in the UV from post-MS stars (mostly from PAGB stars) is slightly larger in 
the far-UV than in the near-UV.
	The proximity of this age estimate to the accepted age of the sun
once again demonstrates the high reliability of the UV-based age estimates 
for intermediate-age, composite populations.
	However, when the new Jimenez models are used, much larger ages
are indicated (10 Gyr giving the best fit).
	This was already evident in Figure 14, where the Jimenez models
appeared much bluer than the Yi models.
	If the continua of G-type stars (including the sun) in the 
Kurucz library are significantly inaccurate, an effort to achieve the right 
age of the sun using population synthesis models and a theoretical spectrum 
would not be appropriate.
	Yet, such tests would still serve as sanity checks for the 
population synthesis computations.

\placefigure{fig17}
\parbox{3.0in}{\epsfxsize=3.0in \epsfbox{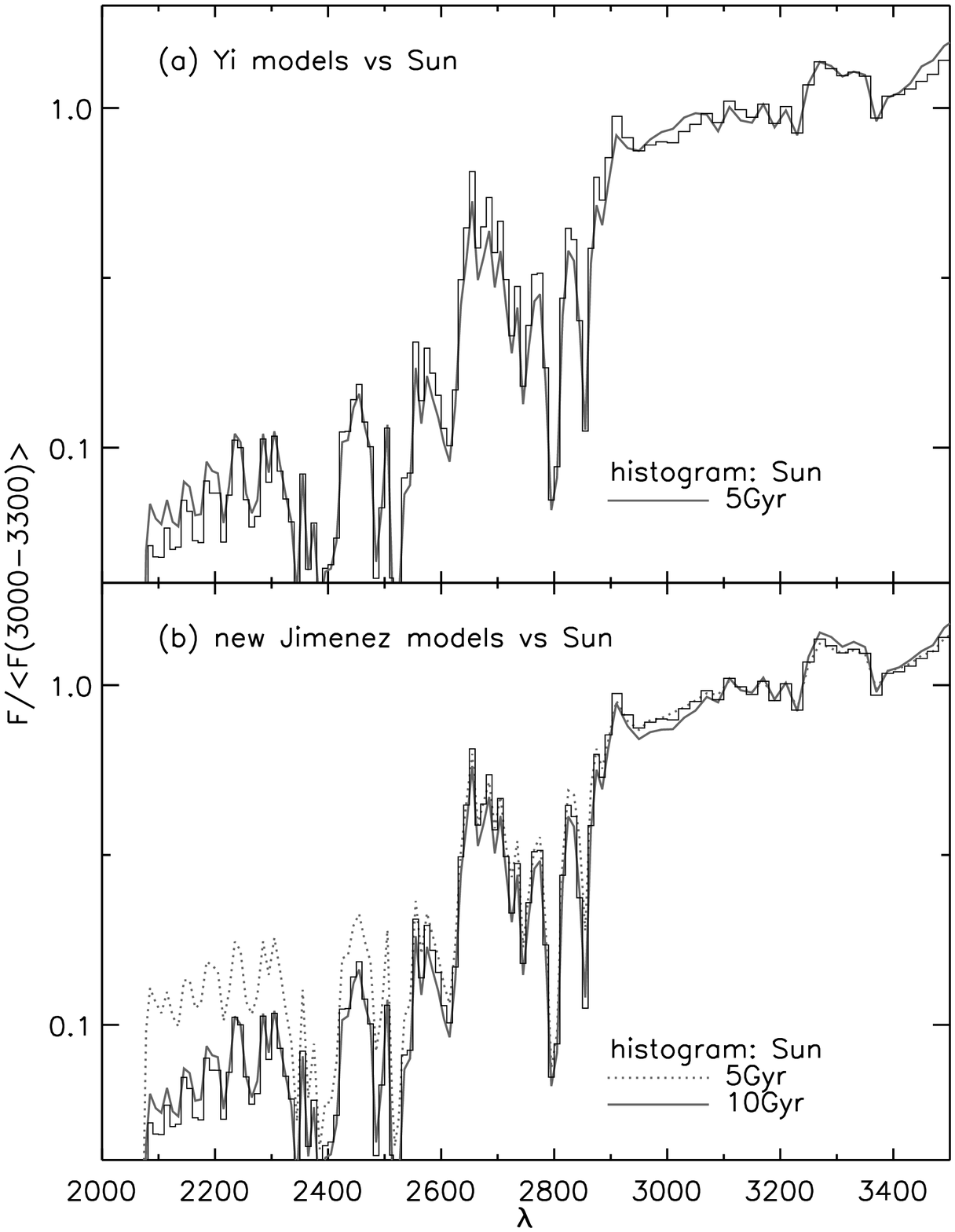}}
\centerline{\parbox{3.0in}{\small {\sc Fig. 17}
A test of the Yi (top) and new Jimenez (bottom) population synthesis models 
on the sun. Models are composite in the sense that they contain all 
evolutionary phases. But, at 5 Gyr, MS stars still dominate the UV spectrum 
(92
should match the solar spectrum at a reasonably close age to the known
one, i.e., 4.5 -- 4.7 Gyr.
}}
\vspace{0.2in}
\addtocounter{figure}{17}

\subsection{Test on M32}

        M32 also provides a good test, as we now 
have resolved visible -- IR CMDs that suggest an age of $\approx 8.5$ Gyr 
(Grillmair et al. 1996). 
	We have adopted the visible -- IR spectrum of M32 from 
Bruzual \& Magris (1997b). 
	Figure 18 shows the fits using the solar abundance models of Yi 
and of Jimenez (new).
	Both the Yi models and the Jimenez models are based on the
theoretical Kurucz spectral library, and thus their fits are limited by 
the known shortcoming of the Kurucz library in matching the spectra of 
cool stars.
	In addition, these single age, single abundance models may not
be good approximations to M32.
	We find that the match becomes better, in particular in the $U$ band
and in the IR, when chemically composite models are used.
	Bearing all these limitations in mind, we cautiously find that the Yi 
models match the M32 spectrum somewhat better.

\placefigure{fig18}

	In conclusion, we note the followings.
	The difference in the age estimate between Spinrad et al. (1997) and
this study is caused by the large difference in the model integrated spectrum.
	The Jimenez models, their preferred models, 
appear much bluer than the Yi and the BC models.
	The new Jimenez models, presumably improved over his earlier version,
are redder than his previous models and thus closer to the Yi and the BC 
models. 
	However, they are still much bluer than the Yi and the BC models.
	Currently, the Yi models are in reasonable agreement with the 
BC models and seem to match the spectra of the sun and M32 at their accepted 
ages better than the new Jimenez models do.

\parbox{3.0in}{\epsfxsize=3.0in \epsfbox{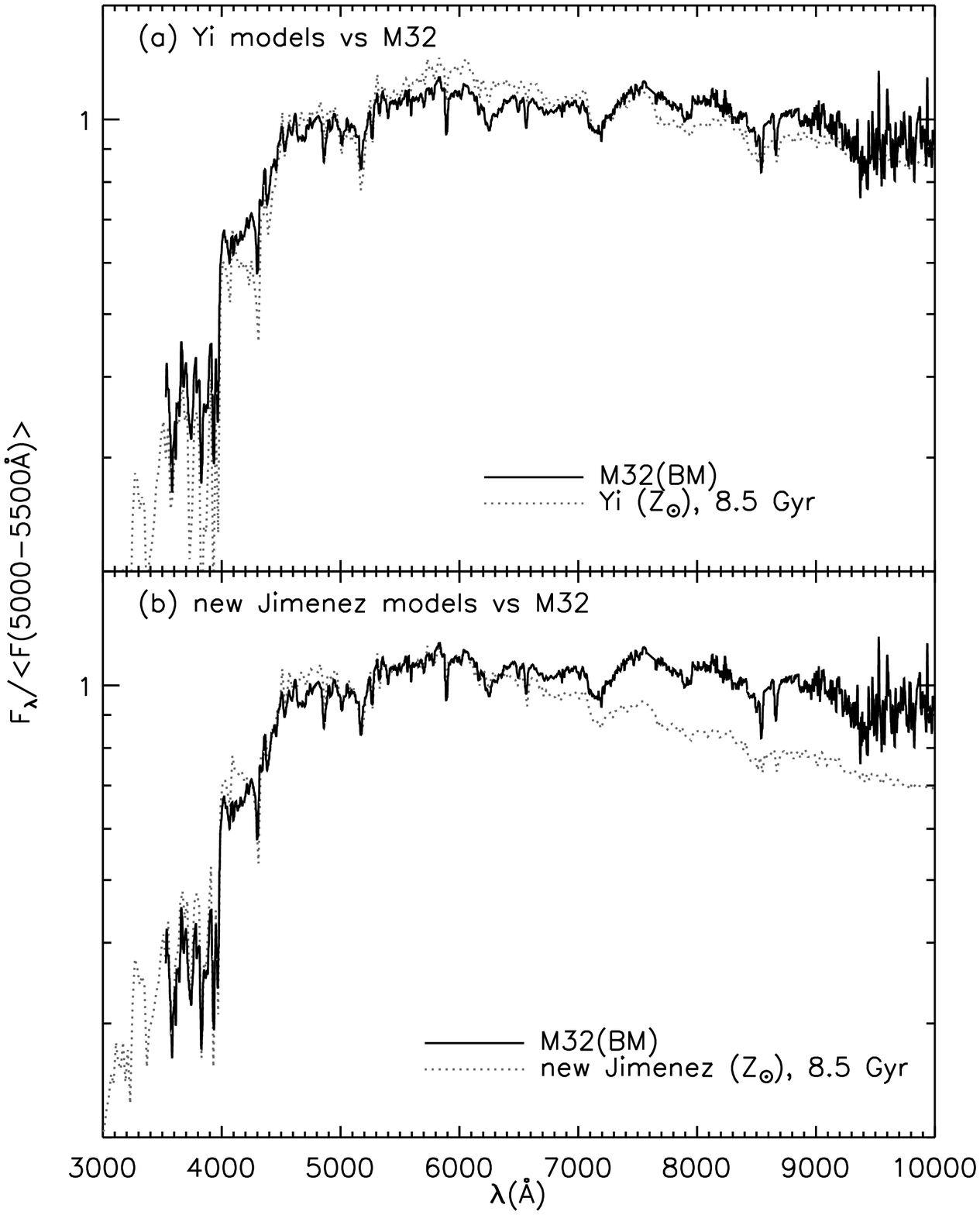}}
\centerline{\parbox{3.0in}{\small {\sc Fig. 18}
A test of the Yi (a) and new Jimenez (b) population synthesis models 
on M32 using its visible -- near IR spectrum from Bruzual \& Margris (1997b). 
Models are for the solar chemical composition and for the age of 8.5 Gyr,
which has been determined from visible -- near IR CMD studies (see text).
The new Jimenez models deviate from the observed spectrum.
}}
\vspace{0.2in}
\addtocounter{figure}{18}

\section{Summary and Conclusion}

	The pioneering studies of Spinrad and his collaborators (Dunlop et al. 
1996; Spinrad et al. 1997) have demonstrated the significance 
of precise age estimates of distant galaxies.
	Their suggestion that the red galaxy, LBDS\,53W091, at $z = 1.552$
is at least 3.5 Gyr old was striking because it would result in a
rather strict constraint on cosmology.
	We have carried out a similar exercise, estimating
the age of this galaxy, but only via continuum fitting.
	When we use the same input parameters, our age estimate is 
approximately 1.4 -- 1.8 Gyr, substantially smaller than theirs, but consistent
with those of Bruzual \& Magris (1997a) and of Heap et al. (1998).

	The large age estimate of Spinrad et al. is apparently caused by the 
use of the early Jimenez models in their analysis, which are significantly 
bluer than the Yi and the BC models.
	The latest Jimenez models are somewhat redder than his earlier models
and closer to the Yi models, resulting in the age estimates that are 
consistent with our estimates.
	This may indicate a resolution of the age discrepancy on LBDS 53W091.

	We have further improved our estimates over previous ones by adopting 
convective core overshoot (OS) and realistic metallicity mixtures.
	The inclusion of OS has little 
effect on the UV-based age estimates, but it raises the age estimates 
based on the visible data normalized to the UV by 20 -- 50\%.
	Adopting realistic metallicity distributions is also important 
because different metallicity groups dominate different parts of the
integrated spectrum.
	If we assume that the majority of stars in LBDS\,53W091 are already 
as metal-rich as those in nearby giant elliptical galaxies, the photometric
data of LBDS\,53W091 indicate up to a factor of two larger ages 
when metallicity mixtures are adopted.

	The UV continua of young galaxies, such as LBDS 53W091, are not
sensitive to OS.
	In addition, solar abundance models reasonably approximate them.
	This relative immunity of the UV data against such complexities
makes the UV spectrum of a distant galaxy a very useful age indicator.
	Our UV-based estimate is approximately $2.0 \pm 0.2$ Gyr and
apparently inconsistent with that of Spinrad et al., 3.5 Gyr, but consistent
with the age estimates we obtained using the new Jimenez models.

	The photometric data of LBDS\,53W091 indicate $1.5 \pm 0.2$ Gyr.
	The slightly larger estimates from the UV continuum fit would be
consistent with this photometry-based one if we include a small amount
of reddening and/or if the core of this galaxy is somewhat older or 
more metal-rich than its outskirt, all of which are quite plausible.
	It may also indicate that there is no substantial age spread among the
stars in LBDS 53W091.

	The age estimates of Spinrad and his collaborators were heavily
based on selected UV spectral breaks.
	This was because UV spectral breaks were believed to be less 
sensitive to the uncertainties in reddening.
	If this is true and if we are ignoring the possible reddening effects, 
our age estimates should be systematically larger than theirs because 
reddening makes the continuum look older, which is opposite to what
we have found.
	Thus the difference between Spinrad et al.'s estimate and ours
cannot be reconciled by adopting any conventional reddening law.

	Our results on LBDS 53W091 are vulnerable to the uncertainties in the 
spectral library in matching the F-type stellar spectra in the UV.
	Such uncertainties may exist not only in the detailed spectral features,
such as those studied by Spinrad et al. (1997) and Heap et al. (1998), 
but also in the continuum.
	The same authors of Heap et al. (1998) are currently obtaining
the UV spectra of F-type stars using HST/STIS.
	When the project is complete, a more accurate analysis both
on the spectral breaks and the continuum will be possible.

	There is no doubt that precise age estimates of high-$z$ galaxies 
would be very useful for constraining cosmology.
	In order to fully take advantage of the power of this technique,
however, we first need to understand the details of the population synthesis,
which are currently creating a substantial disagreement in age estimate.
	We propose to carry out a comprehensive investigation on the various 
population synthesis models through a series of standard tests on the
objects whose ages have been independently determined. 
	Such objects may include the sun, M32, and Galactic globular clusters.
	Our models (the Yi models) currently pass these tests reasonably.

	Our age estimates indicate that LBDS 53W091 formed approximately at
$z = 2$ -- 3.
	However, our smaller age estimate for this {\it one} galaxy does not 
contradict work that suggests galaxies generally formed at high redshifts, 
regardless of the rarity of massive ellipticals at $z \approx 1.5$.  
	Furthermore, we are just beginning to expand our observations 
of galaxies to high redshift, and so the existence of a few old galaxies at 
high redshifts does yet prove any galaxy formation scenario, although it 
can potentially constrain cosmological parameters (in the sense that the 
ages of a few objects can provide lower limits on the age of the Universe 
at that redshift).  
	Finding no old galaxies at high redshift would support a low $z_f$ 
for the general population.  
	Building a larger database of observations is therefore crucial to 
achieve a unique and statistically significant solution.
	Dunlop (1998) reported a discovery of another galaxy (LBDS\,53W069) 
whose UV spectrum looks even redder than that of LBDS\,53W091, although 
its redshift is only slightly smaller ($z=1.43$).
	This would be a stronger sign of large ages of high-$z$ galaxies.
	As more data are collected, our vision to the high-$z$ universe will be
clearer.

\acknowledgements

	This work was encouraged by the open-minded response of Hyron
Spinrad to our initial interest in the work on LBDS\,53W091 done by him 
and his collaborators.
	We thank his group, in particular Daniel Stern, for providing the 
spectrum of LBDS\,53W091.
	We are grateful to Taddy Kodama for providing his metallicity 
distribution models and to Gustavo Bruzual for providing the spectrum of M32.
	The constructive criticisms and comments of Raul Jimenez, 
Hyron Spinrad, Gustavo Bruzual, Sydney Barnes and Pierre Demarque improved 
the manuscript significantly. 
	We owe special thanks to Raul Jimenez and Gustavo Bruzual for making their
models available to us.
	This work was supported by the Creative Research Initiative Program 
of the Korean Ministry of Science \& Technology grant.
	Part of this work was performed while S.Y. held a National Research
Council-(NASA Goddard Space Flight Center) Research Associateship.


{}
\end{document}